\documentclass[12pt]{iopart}

\usepackage{graphicx}  

\begin{document}

\title{Quantum Gravity, Cosmology, (Liouville) Strings and Lorentz Invariance}

\author{Nick E. Mavromatos}
\address{King's College London, Department of Physics,
Theoretical Physics, Strand, London WC2R 2LS, U.K.}

\begin{abstract}I review 
some aspects of non-critical strings in connection with 
Lorentz-Invariance violating approaches to quantum gravity. 
I also argue how non-critical strings may provide a 
unifying framework where string Cosmology and quantum gravity 
may be tackled together.
\end{abstract}

\section{Introduction}

Recently there have been exciting developments in astrophysics 
and cosmology, associated with observations of ultra high energy 
cosmic rays (UHECR), of energies larger than $10^{20}$ eV~\cite{uhecr},
as well as high-redshift ($z \sim 1$) supernovae observations~\cite{snae}
claiming a current-era acceleration of the Universe. The latter observations,
when combined with data from measurements 
of the cosmic microwave background radiation~\cite{cmb}, which provide
evidence for a spatially flat universe, with a $\Omega_{\rm total} =1$, 
seem to indicate that up to $70$ \% of the total energy density of the 
Universe is attributed to a {\it dark energy} component, which could be 
a positive cosmological constant (de Sitter type Universe) 
$\Lambda$ or something else. 
On the other hand, the presence of UHECR events appears puzzling 
from the point of view of Lorentz Invariance~\cite{glashow,kifune,sarkar}. 
The latter symmetry
would impose an energy  threshold 
(Greitsen-Zatsepin-Kuzmin (GZK) cut-off)~\cite{gzk} 
above which certain reactions can occur
which would prevent such UHECR events to reach the observation point,
if one makes the physically plausible assumption that such events come from 
extragalactic sources~\cite{uhecr}.

These data present serious puzzles and challenges 
for the quantum field theory as we know it. 
First, if there is a non-zero cosmological {\it constant} $\Lambda$, 
(de Sitter  phase in the Universe), this would eventually mean that 
the (constant) vacuum energy will dominate over matter in the Universe,
given the $a^{-3}$ relaxation of the latter with the expanding-universe
scale factor $a(t)$. When the vacuum energy is dominant the Universe will enter
again a de Sitter inflationary phase, with exponential growth, $a(t) \sim 
a_0 e^{\sqrt{\frac{\Lambda}{8\pi G_N}}t}$ ($G_N$ is 
the gravitational (Newton's) 
constant). With such scale factors, there will be cosmic horizons, since a light ray in a Robertson-Walker type Universe will take an infinite time 
to traverse a {\it finite distance}
\begin{equation}
\delta \sim \int_{t_0}^\infty \frac{cdt}{a(t)}~<~\infty~, \qquad {\rm for}~
a(t) \sim 
a_0 e^{\sqrt{\frac{\Lambda}{8\pi G_N}}t}
\label{cosmichorizon}
\end{equation} 
The presence of horizons imply the impossibility of defining pure asymptotic
states, which are essential for the proper definition of a 
Heisenberg scattering $S$-matrix, and thus a conventional quantum field theory
in such gravitational backgrounds.  This is a serious challenge also for 
string theory~\cite{challenge}, which, by its very nature, 
is supposed to be a theory of S-matrix. 

Second, a relaxation of the GZK cutoff -in order to explain the UHECR events-
would, in turn, imply a relaxation
of Lorentz invariance~\cite{glashow,kifune}, 
which is another drastic modification
of field theory. Along these lines it has also been 
suggested~\cite{protheroe} that deviation 
from Lorentz symmetry, in the sense of modified dispersion relations 
for matter probes, can also explain the presence of TeV cosmic photon events,
which again should have been prevented from reaching the observation point
within the standard kinematics stemming from Lorentz invariance 
requirement   	
during the scattering 
of energetic photons with the infrared background radiation. 

A question I would like to address in this talk is whether
there is a unified framework in which the above puzzles can be tackled,
with the simultaneous ability to make concrete, 
experimentally falsifiable, predictions. 
In addition, I will attempt to take one more challenge, and 
tackle, within the above framework, 
another important issue, that of the {\it hierarchy} 
between the currently claimed~\cite{snae,cmb} 
``observed value'' of the ``vacuum energy'', $\Lambda/M_P^4 < 10^{-123}$ 
in Planck $M_P$ units,   
and the {\it 
supersymmetry breaking 
scale}, of a few TeV,  in supersymmetric theories. 
For the purposes of my talk I will concentrate in the string theory framework,
and 
present some toy models of (non-critical) strings~\cite{emn} 
where the above issues
can be tackled, as I will argue, in a mathematically consistent way. 
The structure of the talk is the following: 
in section 2 I present a very brief overview of possible violations of 
Lorentz Invariance, and the associated consequences, from various
theoretical viewpoints. 
In section 3  I discuss the issue of Lorentz Invariance in 
string theory. I start from
critical string theory, where Lorentz invariance is valid, 
and then proceed to explain carefully how non-critical
string models may lead to its violation, in the sense of 
modified dispersion relations for matter probes 
(refractive indices {\it etc}). 
In section 4 
I describe briefly two concrete models with the above properties:
(i) one model for space time foam in Liouville strings, 
involving the interaction of string matter with stringy space-time defects,
playing the r\^ole of space time ``foam'', and (ii) 
a non-critical stringy model of Cosmology, where 
the non-criticality, which is here viewed as departure from 
equilibrium, is provided by an initial catastrophic event,
playing the r\^ole of Big Bang. The model consists of two colliding branes.
In the context of this model I will discuss relaxation 
scenaria for the cosmological
vacuum energy, inflation and graceful exit from it, 
and how an S-matrix can be ultimately salvaged by exiting from 
the current-era accelerating phase. 
The model involves two time-like variables, one of which is the Liouville
mode, required for consistency of the non-critical string. 

\section{Lorentz Symmetry Violations and Quantum Gravity: A Brief Overview of Models and Approaches}

I adopt the point of view that Lorentz symmetry is a good symmetry
of any isolated quantum field theory model on flat space time, 
and that 
its possible violations come from either placing the system 
in a `medium' or heat bath (finite temperature), or 
coupling it to (certain models of) quantum gravity, which also 
behave like a stochastic medium~\cite{emn,garay}. 

A theory of quantum gravity must describe Physics at Planck scales, $\ell_p \sim 10^{-35}$ m. At such small length scales the structure 
of space time 
might be very different from what we have experienced so far, {\it e.g.} the 
space time might be discrete, non commutative {\it etc}. 
Therefore Lorentz symmetry
might not be valid  at such small scales,
or at least its form may be different 
from the familiar one characterising the relatively 
low energy scales of the current experiments (for instance, 
the Lorentz symmetry
might be non-linearly realized at Planckian scales).

\begin{figure}[ht]
\begin{center}
\includegraphics[width=.5\textwidth]{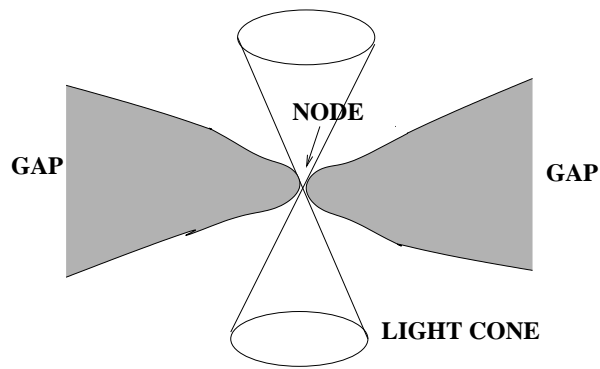}
\end{center} 
\caption[]{ A condensed matter system with nodes in its Fermi surface
(or energy gap). Linearizing the quasiparticle-excitation 
effective theory around 
the nodes one obtains relativistic field theories in the continuum limit.
The `induced light cone' has a limiting velocity which is given by the fermi 
velocity of the node.
Such systems provide analogues of models where Lorentz Symmetry is 
a good symmetry only at low energies.}
\label{fig:nodes}
\end{figure}

There are instructive examples from condensed matter physics, that 
probably help understand physically a possible breakdown of Lorentz symmetry
at higher energies. For instance, 
Antiferromagnetic Spin Systems at low energies
are known to be described effectively by {\it relativistic} continuum 
field theories, for instance the large spin $S \to \infty$ limit of 
planar antiferromagnetic lattice spin systems yields a relativistic $CP^1$ 
$\sigma$-model with action ${\cal S} =\int d^3x \frac{1}{\gamma_0}
|(\partial_\mu - a_\mu)z|^2$, $\gamma_0 \propto 1/S~, S \to \infty$,
$z$ are magnons (spin degrees of freedom), and the non-dynamical gauge field 
$a_\mu$ takes proper account of the correct number of physical degrees
of freedom in the system. 
Another instructive example is that of systems with fermionic quasiparticle
excitations which have {\it nodes} in their fermi surface (or energy gap)
(c.f. figure \ref{fig:nodes}). 
The (average) radius of the fermi surface provides the energy reference  
scale above which all energies of quasiparticle excitations 
are measured. Low energies in this setting are therefore 
viewed those  near the fermi surface. 
Near the node, where the gap function vanishes $\Delta \to 0$, 
the dispersion relation of the quasiparticle excitation
reads: 
$E = \sqrt{\frac{|{\vec k}|^2}{2\mu} + \Delta^2} 
\simeq \frac{1}{\sqrt{2\mu}}|{\vec k}|$, with $\mu$ the Fermi velocity 
of the node, playing here the r\^ole of a
limiting velocity (`speed of light') for the relativistic problem at hand. 
Linearizing the fermionic 
quasiparticle spectrum about the node 
one may therefore obtain 
a relativistic low-energy field theory in the continuum limit 
consisting of Dirac fermions: ${\cal S}_F = \int d^Dx {\overline \Psi}
(\gamma^\mu \partial_\mu + \dots )\Psi $, $\mu=0,1,2, \dots$. 
Such a condensed-matter inspired approach has been adopted
in the literature in an attempt to discuss the origin of Lorentz
symmetry in quantum field theory~\cite{nielsen}.

In the physics of fundamental interactions deviations from Lorentz symmetry 
may become manifest in modifications of the dispersion relations
of matter probes. For massless probes such modifications imply 
non-trivial refractive indices due to the modification in 
the group velocity of the probe. 
For instance, in theories with non-trivial vacuum polarization
at finite temperature, such as quantum electrodynamics {\it etc.},  
one obtains non-trivial refractive indices
already at one loop~\cite{pascual}, in the sense of 
a modified dispersion 
\begin{equation} 
\frac{\partial E}{\partial |{\vec p}|}=
1 + f(T, |{\vec p}|)~, 
\label{convdisp}
\end{equation} 
with $T$ the temperature, and 
the function $f$ being determined from the vacuum  polarization graph.

\begin{figure}[ht]
\begin{center}
\includegraphics[width=.5\textwidth,bbllx=30,bblly=98,bburx=540,bbury=760,clip=]{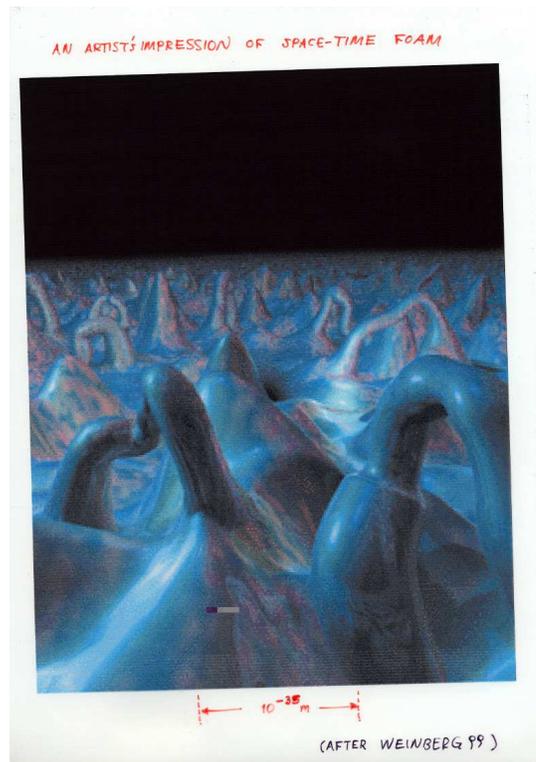}
\end{center} 
\caption[]{An artist impression of space-time foam 
(after S. Weinberg, in Sci. Am. {\it 
``A Unified Physics
by 2050?" (Dec 1999)}; this figure was kindly provided to the author 
by Subir Sarkar).} 
\label{fig:foam}
\end{figure}

It is this sort of modification 
that may be induced also by quantum gravity effects. 
Indeed, one of the most fascinating ideas about quantum 
gravity is the suggestion made by
John Wheeler, and subsequently adopted by Stephen Hawking, 
that space time 
at Planckian scales might acquire a foamy structure 
(c.f. figure \ref{fig:foam}),
which may thus 
behave like a stochastic `medium', with 
non-trivial macroscopic consequences for the propagation 
of matter probes in such backgrounds~\cite{emn,nature,garay}. 
This results in modified
dispersion relation and other non-trivial ``optical'' properties.

Such modified dispersion relations are known to characterise 
non-critical (Liouville) string theory models of space-time foam~\cite{emn}, 
where the departure from criticality (conformality of the associated
$\sigma$-model background) is viewed here as a consequence of the 
interaction of string matter with the (quantum gravity) foam:
\begin{equation} 
E^2 = |{\vec p}|^2 + F(E, |{\vec p}|; \ell_p)~,
\label{gravdisp} 
\end{equation} 
where the function 
$F$ is suppressed by higher powers of the quantum-gravity scale 
(Planck (or string) length $M_P$ ($M_s$)). 
Subsequently, similar phenomena have been argued~\cite{pullin} 
to characterise
the so-called loop gravity approach~\cite{ashtekar}.

The latter is a background independent theory of gravity, 
in which the space time is characterised by a `polymer-like' discrete 
structure. As a result of this property, the basic state of the theory, the ``weavy state'', $|w>$ is such that, although macroscopically 
the (emergent) space time looks Minkowski, nevertheless there is
non-trivial structure at scales smaller than the characteristic 
scale $\ell_W$ where quantum gravity effects set in:
$<w|G_{\mu\nu}|w> = \eta_{\mu\nu} + {\cal O}(\ell_w E)$, where $E$ is the 
energy of the probe. Such effects lead to non-trivial dispersion
relations, similar in nature to those predicted in non-critical strings~\cite{emn}. There is an important difference, though, which distinguishes 
experimentally strings from these theories: 
in loop gravity superluminal propagation
of matter probes is allowed, implying birefringence effects. 
Such superluminal signals are forbidden in the Liouville (non-critical)
string approach to quantum gravity, for specifically stringy 
reasons~\cite{emn}.

There is an important difference of the modification of the dispersion 
relation induced by quantum gravity, as compared to conventional 
vacuum polarization effects~\cite{pascual}, 
in that the former increases with the energy of the probe, in contrast to the conventional field theoretic effects 
which decrease with energy. This allows a possible disentanglement 
of such effects in experiments. One of the most sensitive probes
of such non-trivial optical properties is the observation of light from Gamma
Ray Bursters (GRB)~\cite{nature} or gamma ray flairs~\cite{biller}. 
As mentioned in the introduction, such modified dispersion relations,
have been used subsequently to explain the transparency of the 
Universe in ultra high energy cosmic rays~\cite{kifune} or TeV photon 
events~\cite{protheroe}. 
For a recent and comprehensive review of astrophysical tests of 
such quantum-gravity-induced modified dispersion relations 
see \cite{sarkar}.

Finally, modified dispersion relations for matter probes have been 
recently~\cite{amelino,smolin} argued
to characterise flat space theories in certain models
in which 
the Planck ``length'' is considered a real length. 
As such it should be 
transformed under usual Lorentz boosts. The requirement that the 
Planck length be, along with the speed of light, also 
{\it observer independent} leads quite naturally
to modified ``Lorentz transformations'', and also to modified
dispersion relations for particles. This approach is termed 
``Doubly Special Relativity'' (DSR)~\cite{amelino}. 
The DSR approach should be distinguished 
from the other approaches, mentioned  so far, where  
the Planck length scale is 
associated to a `coupling constant' of the theory, and thus 
is observer independent by definition. 
This is clearly the case of Einstein's General Relativity,
where the Planck scale is related to  
the universal gravitational 
coupling constant (Newton's constant),   
and in string theory (critical and non-critical),
where the Planck length is related to the  
characteristic string scale of the theory, which is also 
independent of inertial observers.

There are two approaches 
so far in the problem of formulating DSR theories: 
in the first~\cite{amelino} 
one postulates a modified dispersion $E^2 = |{\vec p}|^2 + f_1(E, |{\vec p}|; {\tilde \ell}_P)$, where ${\tilde \ell}_P \propto \ell_P$ is the 
inertial-observer independent length, proportional to 
the conventional Planck length $\ell_P \sim 10^{-35}~{\rm m}$. 
The approach is as yet at a 
phenomenological level, and hence one can choose appropriate functions $f_1$.
Let me illustrate the main features with one such choice: $f_1 = -
{\tilde \ell}_P^{-2}\left(e^{{\tilde \ell}_P E} + e^{-{\tilde \ell}_P E} -2\right) + |{\vec p}^2|e^{{\tilde \ell}_PE}$. The generators of the deformed 
Lorentz boosts (assumed, for definiteness and simplicity, to be 
along $z$ direction), 
that leave ${\tilde \ell}_P$ observer independent,
read~\cite{amelino}: 
\begin{eqnarray} 
{\cal N}_z = p_z \frac{\partial}{\partial E} + 
\left(\frac{{\tilde \ell}_P |{\vec p}|^2}{2} + 
\frac{1 - e^{-2{\tilde \ell}_PE}}{2{\tilde \ell}_P}\right)
\frac{\partial}{\partial p_z} - {\tilde \ell}_Pp_z \left(p_j 
\frac{\partial}{\partial p_j}\right)
\label{modboost1}
\end{eqnarray} 
As mentioned previously an important consequence of the formalism is that 
by construction there
are modified dispersion relations:
\begin{equation} 
m^2 = {\tilde \ell}_P^{-2}\left(e^{{\tilde \ell}_P E} 
+ e^{-{\tilde \ell}_P E} -2\right) - |{\vec p}^2|e^{{\tilde \ell}_PE}~,
\label{moddisp1}
\end{equation} 
where $m$ 
is the 
rest mass of a matter point-like probe, defined by the 
appropriate 
quadratic Casimir operator. Note that $m$  
is different from the 
{\it rest energy} ${\cal M}$, 
the latter defined as the  energy at zero momentum ${\vec p}=0$:
\begin{equation} 
m = {\tilde \ell}_P^{-1}\left(e^{{\tilde \ell}_P{\cal M}/2}-
e^{-{\tilde \ell}_P{\cal M}/2}\right)
\label{restmass1}
\end{equation} 
In this particular choice of $f_1$ note that the rest mass differs
from the rest energy only by terms of order ${\tilde \ell}_P^2$ and higher:
$m = {\cal M} + {\cal O}({\tilde \ell}_P^2)$. 

It was remarked in \cite{amelino} that 
the modified dispersion implies a group speed of photon $v_\gamma = 
\frac{\partial E}{\partial p} 
= e^{{\tilde \ell}_P E} = 1 + {\tilde \ell}_P E + \dots $ which diverges as $E \to \infty$. However I note here that 
the meaning of $E \to \infty$ is not clear, especially
in a theory with gravity, where in principle one expects the whole structure
of space time to be changed drastically at energies
close to Planck scale, which thus acts as an effective ultraviolet cutoff. 

In the second approach~\cite{smolin}, one combines Lorentz boosts 
with dilatations in order to achieve the observer-independence
of the Planck length (we assume again 
the boost along $z$-direction for comparison
with the previous approach):
\begin{equation}
{\cal N}'_z = \left[p_z \frac{\partial}{\partial E} \right]
+ E\frac{\partial}{\partial p_z}
- {\tilde \ell}_P p_z \left(E\frac{\partial}{\partial E} + p_j 
\frac{\partial}{\partial p_j}\right)
\label{modboost2}
\end{equation}
where the first term inside square brackets denotes the boost $L_0^z$ , and 
the rest the dilatation ${\cal D}$. 
The combined dilation and boost algebra closes with the three-rotations:
$$ J^i=\epsilon^{ijk}L_{jk}~, K^i \equiv L_0^i + {\tilde \ell}_Pp^i{\cal D}~,
[J^i, K^j]=\epsilon^{ijk}K_k~, [K^i, K^j]=\epsilon^{ijk}J_k~.$$

There is a modified dispersion relation in this approach~\cite{smolin}:
\begin{equation} 
m^2 = \frac{E^2 - |{\vec p}|^2}{(1 - {\tilde \ell}_P E)^2} = 
\frac{\eta^{\mu\nu}p_\mu p_\nu}{(1 - {\tilde \ell}_P E)^2}~.
\label{moddisp2}
\end{equation} 
Notice that in this second approach, in contrast to that of \cite{amelino},
all energies are necessarily smaller than the `gravity scale', $E < 
{\tilde \ell}_P^{-1}$, given that at such scales there appears to be a `metric' collapse in the dispersion relation. This approach is thus closer to the 
conventional understanding of the Planck scale as an ultraviolet cutoff 
for any low-energy theory. Another important feature is that 
the speed of light or, in general, of massless particles is $c(=1)$, as 
in Special Relativity of Einstein.

Again, the rest mass $m $ is different from the rest energy ${\cal M}$, 
but here the difference already occurs already at 
linear order in ${\tilde \ell}_P$:
$m = \frac{{\cal M}}{1- {\tilde \ell}_P {\cal M}} \simeq {\cal M} + {\tilde \ell}_P {\cal M}^2 
+ \dots $.It should be noted that 
there are experimentally testable differences between the 
DSR models of
\cite{amelino} and \cite{smolin}, e.g. in connection with 
kinematical conditions for particle production in collisions~\cite{amelino}.

It remains to be seen what consequences 
these modified flat space transformations
would have in a general relativistic framework. For instance, 
from (\ref{moddisp2}) one would be tempted to identify such 
effects with those in a specific curved  
space-time background, which as I mentioned above,
seems to collapse at Planck energies. Are, then, such DSR models,
when placed in a general relativistic context, selecting 
special metric backgrounds as, for instance, 
is the case of the modified dispersion
relations obtained in non-critical string inspired 
models~\cite{emn}?
These and other questions, related to 
a reconciliation of DSR (viewed as theories
with a minimum length) with quantum mechanics, 
should be pursued in future research.

At this point I end this brief overview of possible theoretical models 
predicting violations or modifications of the Lorentz symmetry. 
In what follows I will concentrate on specific models of such 
violations stemming from (non-critical (Liouville)) String Theory~\cite{emn}.

\section{String Theory and Lorentz Invariance} 

\subsection{Standard Dispersion Relations in Critical Strings} 

Let me start from ``old'' string theory~\cite{green}, living in a 
critical dimension of target space time. From a first quantization viewpoint 
the motion of strings is described by a $\sigma$-model,
a two-dimensional world-sheet field theory, 
in background target-space fields. There is a basic symmetry, called conformal symmetry,
which allows a consistent path integral formulation of 
the $\sigma$-model, and restricts the background fields to their 
target-space classical equations of motion, thereby providing 
an important link between consistent world-sheet quantum geometry 
and target space dynamics. 
It is this symmetry that implies Lorentz invariance in critical 
number of dimensions for flat space times, and the standard dispersion
relations of stringy excitations. 
Let us briefly see how.

Consider a bosonic $\sigma$-model in flat space time, perturbed by certain
tachyon excitations (although tachyons are viewed as instabilities
of bosonic strings, and are absent in superstrings, nevertheless we 
consider them here for simplicity and because they capture all 
the essential features we wish to discuss, which can be immediately generalized to the superstring case):
\begin{equation} 
S_\sigma = \int _\Sigma \partial X^\mu {\overline \partial}X^\nu \eta_{\mu\nu}
+ \frac{1}{4\pi \alpha'}\int d^Dk {\tilde T}(k)\int_\Sigma e^{ik_\mu X^\mu}
\label{tachyon}
\end{equation} 
where $k^\mu$ is a $D$-dimensional target space momentum, 
$\Sigma$ denotes the 
world sheet, $\partial, {\overline \partial}$ are world-sheet derivatives, 
and $\alpha' = \ell_s^2$ is the Regge slope~\cite{green}, related to 
the square of the characteristic length scale of the string $\ell_s$,
which is observer independent from a target-space viewpoint. 
In old string theory this scale was identified with the four dimensional 
Planck mass scale, although in the modern membrane (D-brane)
extension the two scales may not be the same~\cite{membranes},

The exponential vertex operator $V_k=e^{ik_\mu X^\mu}$, which described the 
tachyonic excitation of the bosonic string spectrum, with 
polarization ``tensor'' ${\tilde T}(k)$, is not in general a marginal 
operator in a world-sheet renormalization group sense, and hence under 
the two-dimensional quantum field theory corrections will break the 
conformal invariance in the sense that its conformal dimension 
$\Delta$ is different from one: 
\begin{equation} 
\Delta  = \frac{\alpha'}{2}k_\mu k^\mu 
\label{marginal}
\end{equation} 
Insisting on conformal invariance, i.e. marginality of the operator, 
one obtains the standard dispersion relation ({\it on-shell condition})
for the tachyonic excitation
compatible with Lorentz invariance in target space:
\begin{equation} 
\Delta = 1 \rightarrow  k_\mu k^\mu = \frac{2}{\alpha'} 
\end{equation} 
with the tachyon rest mass $m^2 = -\frac{2}{\alpha'}$ (the negative sign indicates
the abovementioned tachyonic instability of the bosonic string,
but this does not affect the arguments on dispersion relation
because the same procedure applies to all other string excitations).  

The number of target-space dimensions $D$ is also restricted by the 
same requirement of conformal world-sheet invariance if one performs the 
path integral computation of the trace 
of the world-sheet stress tensor $\Theta = T_{\alpha\beta}\gamma^{\alpha\beta}$
in curved world sheets. Even for a free string, then, one finds that 
$<\Theta  > =(D-26)R^{(2)} $, where $<\dots >$ denotes a two-dimensional
path integral average, 
$R^{(2)}$ is the world-sheet
curvature, and the -26 appeared due to the fact that 
world-sheet reprametrization invariance acts from a two-dimensional
viewpoint as a gauge symmetry which needs fixing, thereby implying 
Fadeev-Popov ghosts. Their contribution to the conformal anomaly
is then given by -26$R^{(2)}$. To ensure conformal invariance one must have
$<\Theta>=0$ which select the number of space time dimensions 
for the bosonic string to 26 (for the superstring this number reduces to 10).

\subsection{Non-critical (Liouville) Strings: 
a brief overview of the formalism}

Critical string theory has proven a very successful and elegant 
formalism so far to understand many questions related to the structure
of space time at Planckian distances, to count microstates in 
singular space times such as black holes {\it etc}. 
However, there are some situations which critical strings
appear inadequate to describe. These include formation of black holes 
and other dynamical space-time boundaries, stochastic space-time foam backgrounds (if one adopts the point of view that the latter exist), recoil effects
of stringy solitons, and in general situations which involve the change
of a background over which the string propagates. Critical strings, 
and conventional conformal field theories, describe {\it fixed
background} situations. Since, however, strings are supposed to be a theory
of space time itself, they should be able to provide in principle
an answer as to how the space time was created, and for this 
reason they should exhibit background independence in some sense. 
This is an advantage of the loop gravity approach~\cite{ashtekar} 
which start from abstract fundamental units (``spin networks'') 
and then proceeds to construct a dynamical space time 
out of them. 

\begin{figure}[ht]
\begin{center}
\includegraphics[width=.7\textwidth]{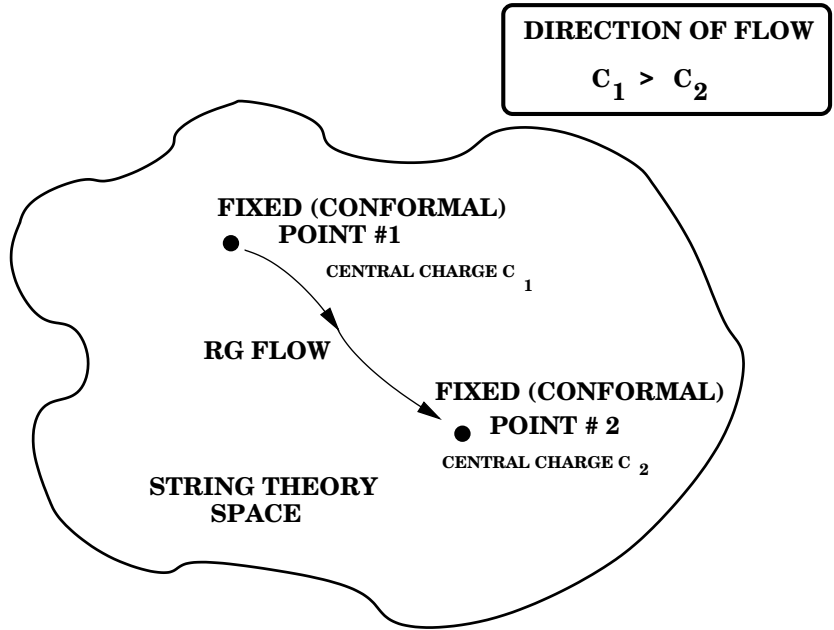}
\end{center} 
\caption[]{A schematic view of string theory space
(which is an infinite dimensional manifold endowed
with a (Zamolodchikov) metric). The dots denote 
conformal string backgrounds. A non-conformal string flows 
(in a tow-dimensional renormalization-group sense) 
from one 
fixed point 
to another (fixed points could even be hypersurfaces in theory space). 
If the string is a unitary field theory 
on the world sheet the direction of the flow 
is towards the fixed point with a lesser value 
of the central charge.}
\label{fig:flow}
\end{figure}

One may argue that such background independence issues 
or background changes in string theory can only be tackled in the context of 
string field theory, which is not developed as yet to a satisfactory 
level of precision or computational power. 
One, however, may be less ambitious and attempt to tackle such issues
perturbatively at first instance, 
within a first quantization $\sigma$-model approach. 
The key to this is probably abandoning {\it conformal invariance}
which fixes the background, and 
use instead {\it non-critical} (Liouville) strings~\cite{ddk},
which is a  mathematically consistent way to deal 
with $\sigma$-models away from their conformal (fixed) points
on the space of string theories (c.f. figure \ref{fig:flow}). 
I will argue below that such a procedure describes consistently
background changes in string theory, and implies, under certain 
circumstances, modified dispersion relations stemming from 
spontaneous breaking of Lorentz symmetry.

To understand better non-critical strings we should first 
mention
a few things about theory space. This is considered
as the space of all possible background target-space-time 
string configurations, which is therefore an infinite dimensional
manifold. The manifold is endowed with a metric, which is 
nothing other than the Zamolodchikov metric of a conformal field 
theory~\cite{zam} given by the appropriate two point function 
of the vertex operators describing the non-conformal deformations:
${\cal G}_{ij}=|z{\bar z}|^2<V_i(z,{\bar z})V_j(0,0)>$, 
where the vertex operators are deforming the $\sigma$-model as:
\begin{equation}\label{deformed} 
S_\sigma = S^* + \int _\Sigma g^i V_i 
\end{equation} 
Here $S^*$ is a fixed-point $\sigma$-model background action,
and $g^i$ are the non-conformal backgrounds/couplings corresponding to 
the vertex operators $V_i$. In stringy models of interest
to us here $g^i=\{ G_{\mu\nu}, \Phi, A_\mu \dots\}$ i.e. is a set
of target-space fields. 

The non-criticality of the deformation is expressed by the 
non-triviality of the renormalization-group (RG) $\beta$-function 
of $g^i$ on the world-sheet: $\beta^i = dg^i/d{\rm ln}\mu \ne 0$,
where $\mu$ is a world-sheet scale provided by the world-sheet 
area~\cite{emn}. Perturbation theory requires that one lies close
to fixed point, which implies that one should work 
with $\beta^i$ expandable in power series
in the couplings $g^i$. Quadratic order is sufficient for our purposes
here, and to this order the $\beta$-function read: $\beta^i = y_ig^i 
+ c^i_{jk}g^jg^k + \dots $, where no sum is implied in the 
first term, $y_i$ is the anomalous dimension, and $c^i_{jk}$ are the 
operator product expansion (OPE) coefficients.

An important property of the {\it off-shell} $\beta^i$ functions 
of stringy $\sigma$-models is their gradient flow form~\cite{zam,emn}
i.e. the fact that they can be derived as $g^i$-space gradients 
of a flow function $C[g,t]$. This function is a renormalization-group invariant
function and coincides with the running central charge of the theory
between fixed points. One has the following relations for its flow along
a renormalization-group trajectory, the celebrated Zamolodchikov 
$c$-theorem~\cite{zam}:
\begin{equation} 
\frac{\partial C[g,t]}{\partial t} = -\frac{1}{12}\beta^i 
{\cal G}_{ij}\beta^j~, \qquad \beta^i = {\cal G}^{ij}\frac{\partial C[g,t]}{\partial g^j}
\label{ctheorem}
\end{equation} 
In unitary $\sigma$-model the metric ${\cal G}_{ij}$  is 
positive definite and hence the flow is such that the $C[g,t]$ 
acts like a thermodynamic $H$ function, decreasing 
{\it monotonically} along the direction 
of the flow. At fixed points 
$\beta^i={\cal G}^{ij}\frac{\partial C[g,t]}{\partial g^j}=0$ and taking into account the positive-definiteness of 
the Zamolodchikov metric one then obtains that the conformal invariance 
conditions are equivalent to target-space equations of motion 
for the flow function $C[g,t]$. The latter, thus, plays the r\^ole 
of a low-energy effective action for the string theory at hand,
which notably includes gravitational interactions. 
At fixed points the flow function becomes the central charge 
characterising the two-dimensional conformal field theory. 
In critical string theories with target space-time interpretation
the flow function becomes 26 for bosonic strings 
(or 10 in superstrings). 

For non-unitary $\sigma$-models, such as stringy $\sigma$-models 
with time-like $X^0$ fields, and time-like dilaton couplings, 
the flow may not be monotonic all the way, but 
one may still 
expect~\cite{kutasov} an overall  
decrease of the running central charge
from one fixed point to the other. This is 
due to the information ``loss'' beyond the ultraviolet cutoff 
of the underlying world-sheet quantum field theory.
In fact the central charge counts degrees of freedom of the system,
and hence there must be an inherent irreversibility if there is 
a cutoff scale. In non-critical strings with time-like fields, 
the running central charge does 
make oscillations before settling to a fixed point~\cite{schmid,emn}.
This will become important later on (section 4) 
when we discuss cosmology in this context.

Let me now argue how non-critical strings become consistent
world-sheet (two-dimensional) 
field theories upon Liouville dressing~\cite{ddk}.
The Liouville mode $\varphi$ seizes to decouple in a non-critical 
string which is characterized by a (RG running) central charge deficit 
$Q^2 \equiv \frac{1}{3}\left(C[g,t] - c^*\right)$ caused by departure
from criticality due to a given deformation (for strings with a 
target-space interpretation $c^*=25$ (9 for superstrings)). 

A detailed analysis~\cite{ddk,emn,schmid} shows that 
the Liouville dressed theory is described by a $\sigma$-model 
action:
\begin{eqnarray} 
&&S_\varphi = Q^2 \int _\Sigma \partial \varphi {\overline \partial} \varphi 
+ \int_\Sigma Q^2 R^{(2)} + \int_\Sigma \lambda^i(\varphi ) V_i ~, \nonumber \\
&& \lambda ^i (\varphi) =g^i e^{\alpha_i Q\varphi} + \frac{\pi}{Q + 2\alpha_i}
c^i_{jk}g^jg^kQ\varphi e^{\alpha_iQ\varphi}~, \nonumber \\
&& \alpha_i \left(\alpha_i + Q \right) = -(\Delta_i - 2)~, {\rm no~sum~over~i}
\label{dressingliouv}
\end{eqnarray}
where $\lambda^i (\varphi)$ are the Liouville dressed couplings and 
the gravitational anomalous dimension $\alpha_i$ is chosen in such a way
so that the conformal dimension of the dressed deformation operator 
$\lambda (\varphi) V_i$ becomes one (marginal). 
Note that this marginality is exact, including interactions ($c^i_{jk}$ 
{\it etc} in the $\beta$-function) 
and 
this constitutes 
the main 
r\^ole of the Liouville dressing, {\it the restoration of 
conformal symmetry}, in such a way that the non-critical, RG flowing 
$\sigma$-model, becomes conformal at the expense of the introduction
of the Liouville mode $\varphi$ in the path integral.

From the quadratic equation that the gravitational anomalous dimensions
satisfy (\ref{dressingliouv}) 
it becomes evident that there are in general two solutions:
\begin{equation}\label{dressings}  
\alpha_i^{\pm} = -\frac{Q}{2} \pm \sqrt{\frac{Q^2}{4} + {\rm sgn}(Q^2 - 25) 
(\Delta_i - 2)}
\end{equation} 
Usually in Liouville theory (central charge $c < 1$) 
one keeps only the $+$ solution, given that 
$\alpha_-$ corresponds to states that ``do not exist'' in the sense of 
not having a well-defined (bounded, normalizable) behaviour in the semiclassical limit where the central charge $|c| \to +\infty$. However in 
supercritical string theory (central charge $c > 25$) this is not the case
and one may, and in some circumstances does, keep both dressings. 
We shall encounter a situation like this in section 4, when we discuss 
specific examples of Liouville strings.

Upon redefining the Liouville mode 
\begin{equation}\label{normal}
\phi \equiv Q \varphi 
\end{equation} 
one may arrive at {\it Liouville-generalized conformal 
invariance conditions}~\cite{ddk}:
\begin{equation} 
{\ddot \lambda}^i + Q(t) {\dot \lambda}^i 
=-{\rm sgn}(C[g,t]-25)\beta^i(\lambda^i(\phi))
\label{liouvcond}
\end{equation} 
where the overdot denotes differentiation with respect to the normalized
Liouville mode $\phi$ (\ref{normal}).

From (\ref{dressingliouv}) it is obvious that the Liouville
mode plays the r\^ole of an extra target-space dimension, which is 
{\it time-like} if the string is supercritical~\cite{aben} 
$Q^2 > 0$ and {\it spacelike} is the string is subcritical. 
It is a general property of Liouville strings that during their flows they 
remain either subcritical or supercritical~\cite{emn,schmid}.
In what follows we shall concentrate on deformations that induce
supercriticality of the string, in which case the Liouville 
mode may be identified with the target time~\cite{emn}.
In fact, what one identifies with the target time is the 
world-sheet zero mode of the normalized field 
\begin{equation}\label{identify}
\phi \equiv Q \varphi \to  t = {\rm target~time}
\end{equation} 
In what follows I shall describe the most important 
physical consequences of such an identification.

\subsection{Absence of S-matrix
and Non-Equilibrium Dynamics 
in non-critical Strings with Time as the Liouville mode}

From (\ref{liouvcond}) it becomes clear that the 
restoration of conformal symmetry leads to 
dynamical equations for the background fields of the non-critical 
$\sigma$-model that have a ``frictional'' form, the damping 
being provided by the (square root of the) central charge deficit $Q^2$. 
Although the sign of $Q^2$ is fixed, however 
the sign of its square root $Q$ is not determined 
in Liouville strings, and as we shall see in section 4,
one may encounter situations~\cite{diamand,gm} 
in which $Q(t)$ evolves in such a way so as to 
change sign (in such a case the string passes through a metastable 
critical point in theory space). 

From (\ref{dressingliouv}) it also becomes clear that 
after the identification (\ref{identify}) 
there will be a dilaton coupling in target space
(defined as the background coupling of the world-sheet curvature term
in the $\sigma$ model action)
$\Phi \sim Q(t)t$ which will contain linear dependences on target time
for $Q$ constant, but in general will be a complicated function of time. 
String theories in such time-like supercritical backgrounds
can be formulated as consistent conformal field 
theories  
only for some special cases of constant $Q$ (linear dilatons)~\cite{aben}
but the corresponding S-matrix aspects are problematic.
One can define factorisable and modular invariant amplitudes ($S$-matrix) 
in such a case 
only for certain {\it discrete} constant values of $Q$, as required
for consistency of the 
mapping of the corresponding conformal theory to a Coulomb gas 
representation with appropriate screening charges at infinity~\cite{aben}.

Nevertheless it is possible to consider more general 
situations, with continuously varying $Q$, provided that one abandons
the concept of a factorisable scattering amplitude~\cite{emn}.
To understand this latter point let us consider  
a generic correlation function 
among $n$ vertex operators
$V_i$ of a Liouville string: 
\begin{eqnarray} 
&&{\cal A}_n = \langle V_{i_1} \dots V_{i_n} \rangle_g = \nonumber \\
&&\int {\cal D}\phi 
 {\cal D}X e^{-S^* -\int _\Sigma d^2\sigma \lambda^i (\phi) 
V_i + \partial \phi 
{\overline \partial} \phi - Q \int_\Sigma d^2\sigma \phi R^{(2)}}
  V_{i_1} \dots V_{i_n}
\label{liouvcor} 
\end{eqnarray} 
where $\phi$ is the normalized 
Liouville mode (\ref{normal}), and $Q^2$ denotes the central charge
deficit, quantifying the departure 
of the non-conformal theory from criticality~\cite{ddk}. 

\begin{figure}[ht]
\begin{center}
\includegraphics[width=.6\textwidth]{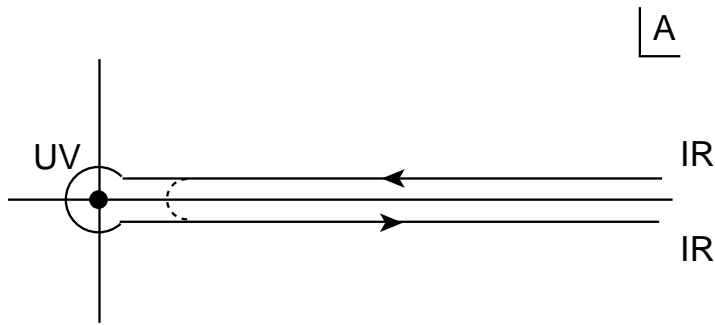}
\end{center} 
\caption[]{Contour
of integration for a proper definition of Liouville field path
integration.
The quantity A denotes the (complex) world-sheet area, 
which is identified with the logarithm of the Liouville 
(world-sheet) zero mode.
This is known in the literature as the Saalschutz contour,
and has been used in
conventional quantum field theory to relate dimensional
regularization to the Bogoliubov-Parasiuk-Hepp-Zimmermann
renormalization method. Upon the interpretation of the
Liouville field as target time, this curve
resembles closed-time-paths in non-equilibrium field theories.}
\label{fig:contour}
\end{figure}

A detailed analysis~\cite{emn} shows that, upon  
performing the world-sheet zero-mode $\phi_0$ integration of the Liouville 
mode $\phi$ in (\ref{liouvcor}), one obtains that the dominant contributions
to the path integral 
are proportional to
\begin{equation} 
{\cal A}_n \propto \Gamma (-s)\langle \dots \rangle '
\end{equation} 
where the prime in $\langle \dots \rangle '$ indicates the absence of 
world-sheet zero mode $\phi_0 \sim {\rm ln}A$ 
of the Liouville field $\phi$, with $A$ the world-sheet area (in units
of the ultraviolet cutoff of the $\sigma$-model), 
$s=\sum_{i} \frac{\alpha_i}{\alpha}
+ \frac{Q}{\alpha}$ and $\alpha_i$ are the Liouville anomalous dimensions
(\ref{dressingliouv}), and $\alpha$ is the anomalous dimension of the 
so-called identity operator $\alpha = -\frac{Q}{2}+ \sqrt{\frac{Q^2}{4} + 8}$. 

In order to compute the integral for arbitrary $s$ (and hence $Q$) 
it is necessary to make an analytic continuation to positive 
integer $s=n^+ \in Z^+$ which 
immediately calls for regularization of  
the prefactors $\Gamma(-s)$. 
This 
can be achieved~\cite{emn,kogan2} by representing the 
integral over the Liouville zero mode by a steepest-descent
contour of $\phi_0$ as indicated in fig. \ref{fig:contour}.
The interpretation of the Liouville zero mode as the target time~\cite{emn}
implies a direct analogy of this contour with {\it closed time like paths}
in {\it non-equilibrium} field theories~\cite{ctp}.

It can be seen~\cite{emn} that the world-sheet ultraviolet limit
of the contour $A \to 0^+$ is ill defined and a regularization is needed
(dashed line in the contour of fig. \ref{fig:contour}). 
This can be inferred by considering infinitesimal Weyl shifts of the 
world-sheet metric (which is integrated over) in the correlator 
(\ref{liouvcor}). Such a divergent behaviour 
implies that the generic 
Liouville correlator (\ref{liouvcor})
is not an S-matrix amplitude, as conventionally assumed in critical strings,
but rather a \$ matrix element connecting asymptotic density 
matrices~\cite{hawk}
$\rho_{\rm out} =$ \$ $\rho_{in}$ with \$ $\ne SS^\dagger$ as a 
result of the world-sheet ultraviolet divergences. 
From a target space time view point small world-sheet areas may be seen as an 
infrared limit of a target space theory, where the strings shrink to their 
point-like limits. Thus, in this picture, the infrared divergences 
in target space are held responsible for the failure of factorisability
of the Liouville string \$ matrix. 

It is this feature that, together with the 
{\it irreversibility} of the world-sheet 
renormalization group flow function, expressed by the 
c-theorem (\ref{ctheorem}),  
makes manifest 
the {\it non equilibrium nature} of non-critical string theory.
There is thus a direct analogy of Liouville strings with open systems
in which there is an entropy change 
(here the entropy is associated with the flow 
function $C[g,t]$~\cite{emn}). 
Further support of the non-equilibrium nature 
comes from the modification of the evolution equation of 
the density matrix of string matter moving in non-critical string backgrounds.
Renormalization-group considerations on the $\sigma$-model theory,
and in particular renormalization-group invariance of physical 
target space quantities, imply the following RG flow equation 
for the density matrix of low-energy string matter $\rho$~\cite{emn,kogan2}: 
\begin{equation}\label{modifiedevol} 
\frac{\partial  \rho}{\partial t} = i[\rho, H ] + :\beta^i {\cal G}_{ij}[g^j, \rho]:
\end{equation} 
where the quantization is achieved in this framework by summing up world-sheet
topologies. In fact it can be shown that canonical quantization in the case of Liouville strings is possible, given that the generalized conformal invariance conditions (\ref{liouvcond}) 
satisfy the Helmholtz conditions for being derived from an off-shell
Lagrangian in $g^i$ space~\cite{emn}. 

Note the similarity of (\ref{modifiedevol}) with the evolution of 
open dynamical systems. Thus we see that the non-criticality of the string,
expressed by the non-vanishing of the world-sheet RG $\beta$-functions, 
leads to the existence of ``environment'' for the subsystem 
of the low-energy string matter. 
In fact an important comment is in order here:
there are quantum ordering ambiguities of the last term due to the fact that
$\beta^i{\cal G}_{ij}$ are complicated power series in the operators
$g^i$ which do not commute with $\rho$. One might 
have thought that these ambiguities may be 
resolved by postulating an ordering that respects formally the 
Lindblad form of the interaction of the string matter with its non-critical
string environment. However, things may not be that simple. 
It must be noted that 
there is a hidden non-linearity in the evolution equation
(\ref{modifiedevol}) due to the string wavefunction dependence
of the Zamolodchikov metric ${\cal G}_{ij} \sim <V_i V_j>$.
When one writes the perturbative expansion $\beta^i{\cal G}_{ij} = \sum_{i_n}
c_{ji_1 \dots i_n}g^{i_1}\dots g^{i_n}$ this dependence
is hidden in the coefficients 
$c_{i_1 \dots i_n} \sim <V_{i_1} \dots V_{in}>_*$ 
where $<\dots>_*$ denote free Liouville string correlators.
Such non-linearities may imply significant deviations from standard
linear Lindblad environmental  entanglement~\cite{szabo}. 
The situation should 
be examined case by case unfortunately, as it seems that general 
statements on this issue cannot be made at this stage.

\subsection{Modified Dispersion Relation in Liouville Strings and Lorentz
Invariance Violation} 

To study the induced modifications on the dispersion relations of string
excitations in the case of non critical deformations 
we concentrate on the generic Liouville dressing (\ref{dressingliouv}).
Upon the identification of the Liouville mode with time (\ref{identify})
we may rewrite this relation as:
\begin{equation} 
\lambda^i(\phi) = \lambda^i(t) e^{i\left(\alpha_i + \Delta \alpha_i \right)t}
\end{equation} 
where $\Delta \alpha_i$ depends on the OPE coefficients
$c^i_{jk}$ which encode the non-critical interactions
of the string probe with the ``environment''. 
In string theory the OPE, which are related to string target-space 
scattering amplitudes, are power series in $\alpha ' |{\vec k}|^2$
for closed strings, and $\sqrt{\alpha} |{\vec k}|$ for open strings.
Let us concentrate in the case of closed strings for definiteness,
and also for physical concreteness, given that closed strings
include gravity. We restrict ourselves to {\it massless modes} for 
our purposes below. 
To order $g^2$ (i.e. close to a fixed point), 
where we restrict ourselves for our purposes 
in this article, the  OPE are assumed of order $\alpha ' |{\vec k}|^2$. 
On the other hand, for vertex operators proportional to $e^{ik_\mu X^\mu}$
(in Fourier space), describing the string excitation spectrum in
flat space times,  the anomalous dimensions $\alpha_i \sim  |{\vec k}|^2$.
This can be found from (\ref{dressings})
by noting that the central charge deficit $Q^2$ is 
determined from the c-flow theorem (\ref{ctheorem})
$Q^2(t) \sim  \int^t \beta^i {\cal G}_{ij} \beta^j ={\cal O}(|{\vec k}|^4)$,
since $\beta^i ={\cal O}(|{\vec k}|)$ for reasons stated previously. 

Thus, for massless closed string excitations one has the estimate:
\begin{equation} 
\alpha_k + \Delta\alpha_k \sim \ell_s |{\vec k}| 
\left( 1 + {\cal O}(|{\vec k}|^2\ell_s^2)\right)^{1/2}
=\ell_s |{\vec k}| + {\cal O}(\ell_s^3 |{\vec k}|^3)
\end{equation} 
The complete Liouville-dressed vertex operator, therefore,
describing the massless string excitation reads:
\begin{equation} 
\lambda^k(t)V_k \sim g^k e^{i(\alpha_k + \Delta \alpha_k)t}
e^{i{\vec k}\cdot {\vec X}} \sim e^{i(|{\vec k}| + {\cal O}(\ell_s^2|{\vec k}|^2))t + i{\vec k}\cdot {\vec X}}
\label{fullvertex}
\end{equation} 
The coefficient of the Liouville mode (time) is defined as the energy 
of the probe, and hence from (\ref{fullvertex}) 
one obtains a modified dispersion relation for massless string 
excitations~\cite{emn}:
\begin{equation}\label{modifiedstring}
E \sim |{\vec k}| + \eta \ell_s|{\vec k}|^2
\end{equation} 
where the coefficient $\eta$ and its sign is to be determined 
in specific models of non-critical strings, and the string scale
$\ell_s = \sqrt{\alpha'} $ may be taken to be the Planck scale 
$M_P^{-1}$ of 
quantum gravity (but, we repeat again, in modern versions of
string theory where our world is viewed as a string 
membrane~\cite{membranes} 
$\ell_s$  is an arbitrary free parameter,
which is in general different from the four-dimensional 
Planck scale $M_P^{-1}$). 

From the modified dispersion one obtains a refractive index for 
the massless probe in the sense of its group velocity not 
being equal to one:
\begin{equation} 
\frac{\partial E}{\partial |{\vec k}|} = 1 + 2\eta \ell_s|{\vec k}| + \dots 
\label{refractive}
\end{equation} 
which implies violation of Lorentz invariance. 
The latter should not come as a surprise given the non-equilibrium
stochastic nature of the non-critical string (c.f. (\ref{modifiedevol}))
which resembles an open system in which string matter propagates
in a `stochastic' dissipative way (c.f. (\ref{liouvcond})). 

However, in our approach~\cite{emn} 
the Lorentz violation is {\it spontaneous} in the sense that 
the full string 
theory may be critical, and the non-criticality is only a result of 
restricting oneself in an effective low-energy theory (``ground state'' 
not respecting the symmetry). 
In this last sense the presence of the Liouville mode is a {\it collective}
description of ``environmental'' effects associated with 
degrees of freedom 
essentially 
unobserved
by a local observer~\cite{emn}. 

The precise nature of such degrees of freedom can be unveiled   
in realistic concrete examples of non-critical strings with physical 
significance. We examine two such examples in the next section. 

\section{Concrete Physical Examples of Liouville Strings} 

\subsection{A Liouville string model for Space Time foam}

\begin{figure}[ht]
\begin{center}
\includegraphics[width=.6\textwidth]{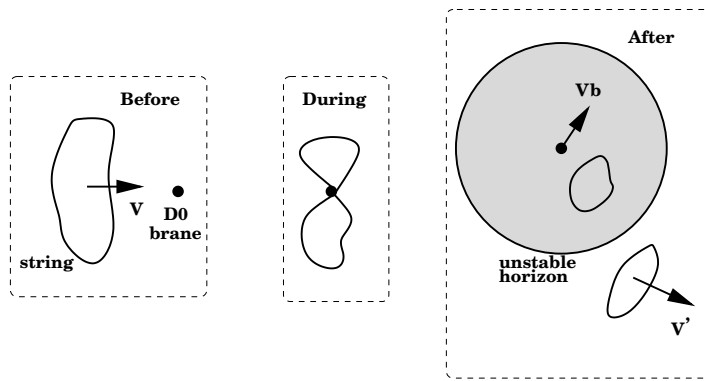}
\end{center} 
\caption[]{A schematic representation of D-particle foam; closed strings
scatter off D0 brane defects in space time. The scattering 
distorts the surrounding 
space time, causing the formation of unstable horizons with non-trivial 
refractive index in their interior. These structures capture part of 
the propagating string matter, thereby leading to stochastic medium behaviour,
and hence a space-time foam situation.}
\label{fig:dfoam}
\end{figure}

We describe below a toy model for stringy space-time foam~\cite{emn}, 
consisting 
of a bulk space time ``punctured'' with D(irichlet)-0-brane point-like defects.
The latter are solitonic states in string theory~\cite{membranes}, and
as such they are not simply structureless 
point-like objects, but have substructure in 
the sense of an infinity of internal degrees of freedom corresponding 
to the massive string states. Once a closed string, representing 
light string matter in the model,  is scattered off the defects,
the latter recoil and distort the surrounding space time. 
The distortion is expressed through the dynamical formation of (unstable)
horizons, the interior of which is found to have 
a different refractive index from the exterior flat Minkowski space time.
The situation is schematically depicted in figure \ref{fig:dfoam}.

The mathematical formalism is based on 
a conformal field theory treatment of recoil in D-brane theory~\cite{kmw}
which allows for changes of $\sigma$-model backgrounds. This can be
achieved by the so-called 
logarithmic conformal field theories
(LCFT), 
which lie on the border line 
between conventional conformal field theories and general two-dimensional
field theories, and can still be classified by conformal data. 
Below we review briefly the main results.

A 
D-particle, recoiling with velocity $v^i$, and initially positioned at $y^i$, 
can be viewed as a D-particle 
with open string excitations attached to it (c.f. middle 
figure \ref{fig:dfoam}), which from a $\sigma$-model point of view are described by adding to the free string $\sigma$-model the following 
boundary (world-sheet disk)
deformations
\begin{equation} 
{\cal V}_{\rm rec,rest} = \int_{\partial \Sigma} \left(\epsilon^2 y_i \theta_\epsilon (X^0)\partial_nX^i + 
 \epsilon v_i X^0 \theta_\epsilon (X^0) \partial_n X^i \right)
\label{recoil} 
\end{equation} 
where $\partial_n $ is a normal world-sheet boundary derivative, 
and we assumed that the D-particle was initially {\it at rest}. 
These operators form a LCFT pair of operators~\cite{kmw}.
In the above formula $\theta_\epsilon (X^0) \sim \theta(X^0) e^{-\epsilon X^0}$
for $X^0 >0$ is a regularized Heaviside operator, with $\theta(X^0)$ the 
ordinary Heaviside function. The coordinate $X^0$ obey Neumann 
boundary conditions on the world-sheet boundary $\partial \Sigma$, whilst
$X^i$, $i$ spatial index, obey Dirichlet boundary conditions, and express
the fact that the D-particle defect can trap strings with their ends 
attached on it (c.f. figure \ref{fig:dfoam}). In the limit $\epsilon \to 0^+$ 
which we consider here the first term in (\ref{recoil}) 
is subleading and from now on we ignore it. 

We next extend the discussion to motion of string matter 
through a gas of moving D-particles~\cite{egmn}, 
as is likely to be the case for
a laboratory on Earth, e.g., if the D-particle foam is comoving 
with the Cosmic Microwave Background (CMB) frame. 
Assuming this to be moving with three-velocity ${\vec w}$ 
relative to the observer,
the recoil deformation takes the following form
to leading order in $\epsilon \to 0^+$:
\begin{eqnarray}
{\cal V'}_{\rm rec} 
= \int _{\partial \Sigma} 
\theta_\epsilon (-X^0_w)\gamma_w w^i\partial_{\rm n} X^i_w
+  \int _{\partial \Sigma} 
\theta_\epsilon(X^0_w)
\gamma_w (w^i+ \epsilon v^i(w))\partial_{\rm n} X^i_w
\label{changerec}
\end{eqnarray}
where the suffix $w$ denotes quantities 
in the boosted frame. The recoil velocity $v^i(w)$ 
depends in general on $w$, and is determined by momentum
conservation during the scattering process, as discussed in~\cite{kmw}.
The main novelty in the $w \ne 0$ case 
is that now there are two $\sigma$-model operators in (\ref{changerec}).

The deformations (\ref{recoil}) and (\ref{changerec}) are relevant 
world-sheet deformations in a two-dimensional renormalization-group 
sense, with anomalous scaling dimensions $-\epsilon^2 /2$~\cite{kmw}. 
Their presence drives the stringy $\sigma$-model non-critical,
and requires dressing with the Liouville mode $\phi$. 
As mentioned previously, in section 3,  
in Liouville strings there are two screening operators 
$e^{\alpha_\pm \phi}$, where the 
$\alpha_\pm$ are the Liouville-string anomalous dimensions
given by:
\begin{equation} 
\alpha_\pm = -{Q \over 2} \pm \sqrt{\frac{Q^2}{2} + \frac{\epsilon^2}{2}}
\end{equation} 
and the central charge deficit $Q$ was computed in~\cite{emn} 
with the help of the c-theorem (\ref{ctheorem}), 
and 
found to be of higher order than $\epsilon^2$. Hence
$\alpha_\pm \sim \pm \epsilon$. 

The $\alpha_-$ screening operator is sometimes neglected because it
corresponds to states that do not exist in Liouville theory. However, this
is not the case in string theory, where one should keep both screenings as
above. This is essential for recovering the correct vanishing-recoil limit
in the case of infinite D-particle mass.
The Liouville-dressed boosted deformation then reads: 
\begin{eqnarray}
&&{\cal V'}^L_{\rm rec} =
\int _{\partial \Sigma}e^{\alpha_-\phi}  
\theta_\epsilon (-X^0_w)\gamma_w w^i\partial_{\rm n} X^i_w
+  \nonumber \\
&& \int _{\partial \Sigma} e^{\alpha_+\phi}  
\theta_\epsilon(X^0_w)
\gamma_w (w^i+ \epsilon v^i(w))\partial_{\rm n} X^i_w
\label{liouvdressed}
\end{eqnarray}
Using Stokes' theorem, and ignoring terms that vanish
using the world-sheet equations of motion, one arrives easily at
the following bulk world-sheet operator:
\begin{eqnarray} 
{\cal V'}^L_{\rm rec} =
\int _{\Sigma}e^{\epsilon \phi}  
\theta_\epsilon (X^0_w)\epsilon^2  
\gamma_w v^i(w)\partial_{a} X^i_w\partial^a \phi 
+ \dots,~a=1,2,  
\label{liouvdressedbulk}
\end{eqnarray}
where the $\dots $ denote terms subleading as $\epsilon \to 0^+$ as we shall
explain below. Notice the cancellation of the terms proportional to 
$w$ due to the opposite screening dressings. Recalling that the 
regularised Heaviside
operator $\theta_\epsilon (X) = \theta_0 (X) e^{-\epsilon X}$, where
$\theta_0 (X)$ is the standard Heaviside function, we observe that one can
identify the {\it boosted time coordinate} $X_w^0$ with the Liouville mode
$\phi$: $\phi = X_w^0$~\cite{emn}. At long times after the 
scattering, the Liouville-dressed theory leads to 
target-space metric deformations of the following form to order 
$\epsilon^2$: 
\begin{equation}\label{metricdressed} 
G_{\phi i} = 
\epsilon^2 \gamma_w v^i(w) \phi
\end{equation}
As explained in detail in~\cite{kmw,emn}, at the long times 
after the scattering event when the $\sigma$-model formalism
is valid, one has $\epsilon^2 \phi_0 \sim 1$: $\epsilon$ and
the world-sheet zero mode of $\phi$, $\phi_0$,  are not independent
variables, as $\epsilon$ is linked
with the world-sheet renormalization-group scale.  

The Lorentz 
breaking effects are proportional to $v^i(w)$,
which is determined by momentum conservation
and hence is independent of $w$ in the non-relativistic limit.
In the case of photon scattering
off D-particles, one would have $v^i(w) =\sqrt{(1+ |w|)/(1-|w|)}v^i(0)$. 
{\it Our form of Lorentz violation exhibits Galilean 
invariance} for small $w$. This  
is an important feature which differentiates
our Liouville foam model from generic Lorentz-violating
models of space time foam considered in the literature~\cite{SUV}.  
For our purposes, the main effect of our D-particle model
for space-time foam is the modification
of the dispersion relation for an energetic particle, which causes it to
propagate more slowly than a less energetic particle.

The formation of horizons, mentioned earlier, and 
depicted in figure \ref{fig:dfoam}, is obtained 
upon considering 
quantum fluctuation effects, 
which express
the probability of finding a particular defect configuration as a
fluctuation around a mean value. 
Formally, one can evaluate quantum
fluctuations in the recoil process by making an appropriate resummation
over the world-sheet genera, which lead to stochastic 
probability distribution
functions in the $\sigma$-model path integral~\cite{kmw}.
Performing appropriate coordinate transformations, which from a
$\sigma$-model view point amount to redefining path integral variables,
and averaging over quantum fluctuations 
one can map the metric (\ref{metricdressed}) in the limit $w=0$
into one which has an
expanding horizon surrounding the defect at:  $r_{\rm horizon}^2 =
t^2/{b'(E)}^2$, with ${b'(E)}^2 =
4g_s^2 \left(1 - \frac{255}{18}g_s^2\frac{E_{\rm kin}}{M_D}\right)
+ {\cal O}(g_s^6)$, 
$E_{\rm kin} (\equiv E)$ is the kinetic recoil energy of the 
non-relativistic D-particle, 
$g_s < 1$ is the string coupling, assumed weak for the 
validity of the $\sigma$-model perturbative expansion, and 
$M_D=M_s/g_s$ is the D-brane mass scale with $M_s$ the string scale.
Often one assumes $M_s$ close to the four dimensional Planck mass
of $M_P \sim 10^{19}$ GeV, although in general $M_s, g_s$ are 
parameters which can be 
constrained by data~\cite{nature}. 
Inside the horizon, the space-time metric has the form:

\begin{equation}
\label{metricrecoil}  
ds^2_{\rm inside} = \frac{{b'(E)}^2 r^2}{t^2}dt^2 - \sum_{i=1}^{3}dx_i^2~:
\qquad r^2 =\sum_{i=1}^{3} x_i^2,
\end{equation}
and one can match this space-time with a flat external Minkowski
space-time, as explained in detail in~\cite{emn}. The internal
`bubble' metric configuration (\ref{metricrecoil}) violates the positive
energy conditions and hence is {\it unstable}.  Thus a `breathing bubble'
picture emerges from such a world-sheet path integration of the Liouville
zero mode~\cite{emn}, in which the average lifetime of the bubble is
Planckian, during which the bubble expands to its maximum size, which is
of order the Planck length, and then recontracts.

We have identified three possible experimental implications of the
formation of such quantum space-time bubbles. {\it First}, low-energy
particles may get trapped inside the bubble, and their absorption within
the space-time foam provides an explicit mechanism for the apparent loss
of information accessible to external observers. {\it Secondly}, in the
case of electrically-charged particles, their non-uniform (spiral) motion
inside the bubble causes the emission of radiation, which may escape in
the form of photons. Since the bubble interior has a non-trivial
refractive index~\cite{emn}, these particles exhibit transition
radiation, i.e., the emission of photons accompanying an
electrically-charged particle when it crosses an interface separating two
media with different refractive indices. A fraction of this radiation
escapes from the bubble, yielding photons that accompany the charged
particle. {\it Thirdly}, propagation inside the bubble causes the particle
to travel more slowly, as seen by an external observer. This effect is
equivalent to the retardation induced by the modification
(\ref{metricdressed}) of the space-time metric, as the D-brane defect recoils
when struck by the propagating particle.

The refractive index in this D-brane foam can be computed as~\cite{emn} 
\begin{equation}
v=1 - {\cal O}(\frac{g_sE}{M_s})
\label{branerefr}
\end{equation} 
where $E$ is the kinetic energy transfer of the matter 
probe during the collision
with the D-particle defect. 
The index (\ref{branerefr}) is always {\it subluminal}, thereby 
differentiating the model from certain models of loop gravity
where superluminal propagation also occurs~\cite{pullin}. 
The subluminal nature
is due to specific stringy properties of the model, in particular 
it stems from the fact
that the effective action describing the recoil excitations is of 
Born-Infeld form of non-linear ``electrodynamics''~\cite{kmw}, 
and hence there is 
a limiting (light) velocity for signal propagation. 

For astrophysical and other tests of these string-inspired dispersion 
relation see \cite{sarkar}. We should mention, though, that the non-critical
string model of space time foam of \cite{emn},
despite its minimal Planck-scale suppression of the effects, 
seems to avoid severe
constraints, 
especially from atomic physics experiments~\cite{SUV,egmn}, 
which seem to exclude  
the loop-gravity minimal suppression models. 
Thus, for this model of foam, 
GRB arrival-time tests~\cite{nature} seem, as yet, to provide
the most sensitive probes. 

\subsection{A Liouville String approach to Colliding Branes Cosmology: 
Inflation, and Supersymmetry-Breaking/Vacuum-Energy Hierarchy} 

As mentioned in the introduction, 
there has been recent exciting evidence from astrophysical 
observations on high redshift type I Supernovae~\cite{snae}
that our Universe is currently accelerating. 
The evidence has been seconded by cosmic microwave background data~\cite{cmb}
pointing towards total flatness of our Universe $\Omega_{\rm total}=1$,
a result which, 
when combined with the estimate for the total matter 
contribution (including dark matter) $\Omega_{M} \sim 0.3$,
implies that there is a dark energy component in our Universe
which is 
70 \% of the total energy density. This is in agreement with 
the Supernovae
data implying acceleration of the Universe, and implies that
the Universe is currently in a phase where this dark energy component 
begins to take over. These data are in agreement with 
naive cosmological constant in the Universe (de Sitter), 
but this issue
present an enormous theoretical challenge, due to the cosmic horizon problem
(\ref{cosmichorizon}) and its incompatibility with the definition
of a scattering matrix~\cite{challenge}, and hence string theory. 

As we have seen in section 3, Liouville non-equilibrium strings, 
are not characterised
by a proper S-matrix, nevertheless they could be quantized consistently
on the world-sheet. It is therefore legitimate to argue~\cite{emnaccel}
that, if our Universe has a cosmological constant, then the only way to 
quantize it within string theory is to postulate some sort of 
Liouville string cosmology. 
One, however, may actually attempt to go one step further and
try to find cosmological non-critical string models  which eventually
asymptote (in time) to equilibrium (critical string) cosmologies.
In such models, which resemble {\it quintessence models}~\cite{carroll}
in conventional cosmology, there is an initial cause for departure
from criticality (conformality of the respective $\sigma$-model,
i.e. the presence of a central-charge deficit $Q^2 >0$),
which imply that the Universe 
undergoes a non-equilibrium, and then a relaxation 
phase, which lasts until today. 
Eventually the central charge deficit vanishes, and the string theory
reaches its critical equilibrium situation. This allows an appropriate
definition of asymptotic states and an $S$-matrix, since in such models one eventually exits from the de Sitter accelerating phase. 

In what follows I will give a concrete example of such a non-equilibrium
cosmology. The model involves 
two colliding branes worlds, one of which is assumed to be our 
world. We shall be very brief in our exposition. For details we refer the 
reader to \cite{gm} where the model is described in some detail. 
I will argue below that the model produces inflation 
and a relaxing to zero cosmological ``constant''
hierarchically small as compared to the supersymmetry breaking (TeV) 
scale. Supersymmetry breaking is induced by compactification 
of the brane worlds on magnetized tori. 
The crucial ingredient is the non-criticality (non conformality) 
of string theory
on the observable brane world induced at the collision, which is thus
viewed as a cause for departure from equilibrium in this system.
The hierarchical smallness of the present-era vacuum energy, 
as compared to the SUSY breaking scale, is attributed to relaxation
phenomena.

The model 
consists of two five-branes of type IIB string theory, 
embedded in a ten dimensional bulk space time. 
Two of the longitudinal brane dimensions are assumed compactified
on a small torus, of 
radius $R$. In one of the branes,
from now on called {\it hidden}, the torus is {\it magnetized} 
with a 
constant magnetic
field of intensity H. This amounts to an effective 
four-dimensional vacuum energy contribution in that brane of order:
$R^2H^2 > 0$. Notice that such compactifications
provide alternative ways of breaking supersymmetry~\cite{bachas},
which we shall make use of in the current article. 
In scenaria with two branes embedded 
in higher-dimensional bulk space times it 
is natural to assume (from the point of view
of solutions to bulk field equations) that the two branes
have {\it opposite} tensions. We  assume that 
before the collision the visible 
brane (our world) has positive tension $V_{\rm vis}=-V_{\rm hidd} > 0$.
The presence of opposite tension branes implies that 
the system is not stable, but this is O.K. from a cosmological
view point. 

For our purposes we assume that the two 
branes are originally on {\it collision course}
in the bulk, with a relative velocity $u \ll 1$ for the 
validity of the $\sigma$-model perturbation theory, and to allow
for the model to have predictive power. 
The collision takes place at a given time moment. 
This constitutes an event, which 
in our scenario is identified with the {\it initial
cosmological singularity} (big bang) on the
observable world. We note that 
similar scenaria
exist in the so-called ekpyrotic model for the Universe~\cite{ekpyrotic}.
It must be stressed, though, that 
the similarity pertains only to the brane-collision event. 
In our approach the physics is entirely different from 
the ekpyrotic scenario. The collision  
is viewed as an event resulting in non-criticality 
(departure from conformal invariance) of the underlying string theory,
and hence in non-vanishing $\beta$ functions at a $\sigma$-model level. 
On the contrary, in the scenario of \cite{ekpyrotic} the 
underlying four-dimensional effective theory (obtained after integration
of the bulk extra dimensions~\cite{ekpyrotic,linde}) is assumed always
critical, satisfying classical equations of motion, and hence vanishing 
$\sigma$-model $\beta$
functions. This latter property leads only to contracting 
and {\it not expanding} four-dimensional
Universes according to the work of \cite{linde}, which constitutes
one of the main criticisms of the ekpyrotic universe. 
On the other hand, in our non-critical description of the collision
we do not assume classical solutions of the equations of motion,
neither specific potentials associated with bulk branes, as 
in \cite{ekpyrotic}. 

The physics of our colliding worlds model can be summarized as follows: 
During the collision one assumes 
electric current transfer 
from the hidden to the visible brane, which results in 
the appearance of a magnetic field on the 
visible brane. We also assume that the entire effect
is happening very
slowly and amounts to a slow flow of energy and current
density from the hidden to
the visible brane (our world). 
In turn, this results in a
positive energy component
of order $H^2R^2$  
in the vacuum energy of the visible brane world.

At the moment of the collision the conformal invariance 
of the $\sigma$-model describing (stringy) excitations on the observable
brane world is spoiled, thereby implying the need for Liouville 
dressing~\cite{ddk,emn}. This procedure restores 
conformal invariance at the cost of introducing 
an extra target space coordinate (the Liouville mode $\phi$),
which in our model has time-like signature.
Hence, initially, one
faces a two-times situation. 
We argued~\cite{gm}, though, that
our observable (cosmological) time $X^0$ parametrizes a certain curve, 
$\phi={\rm const.}\;X_0+{\rm const.'}$, on the
two-times plane $(X_0,\phi)$, and hence one is left with one
physical time.  

\begin{figure}[ht]
\begin{center}
\includegraphics[width=.8\textwidth]{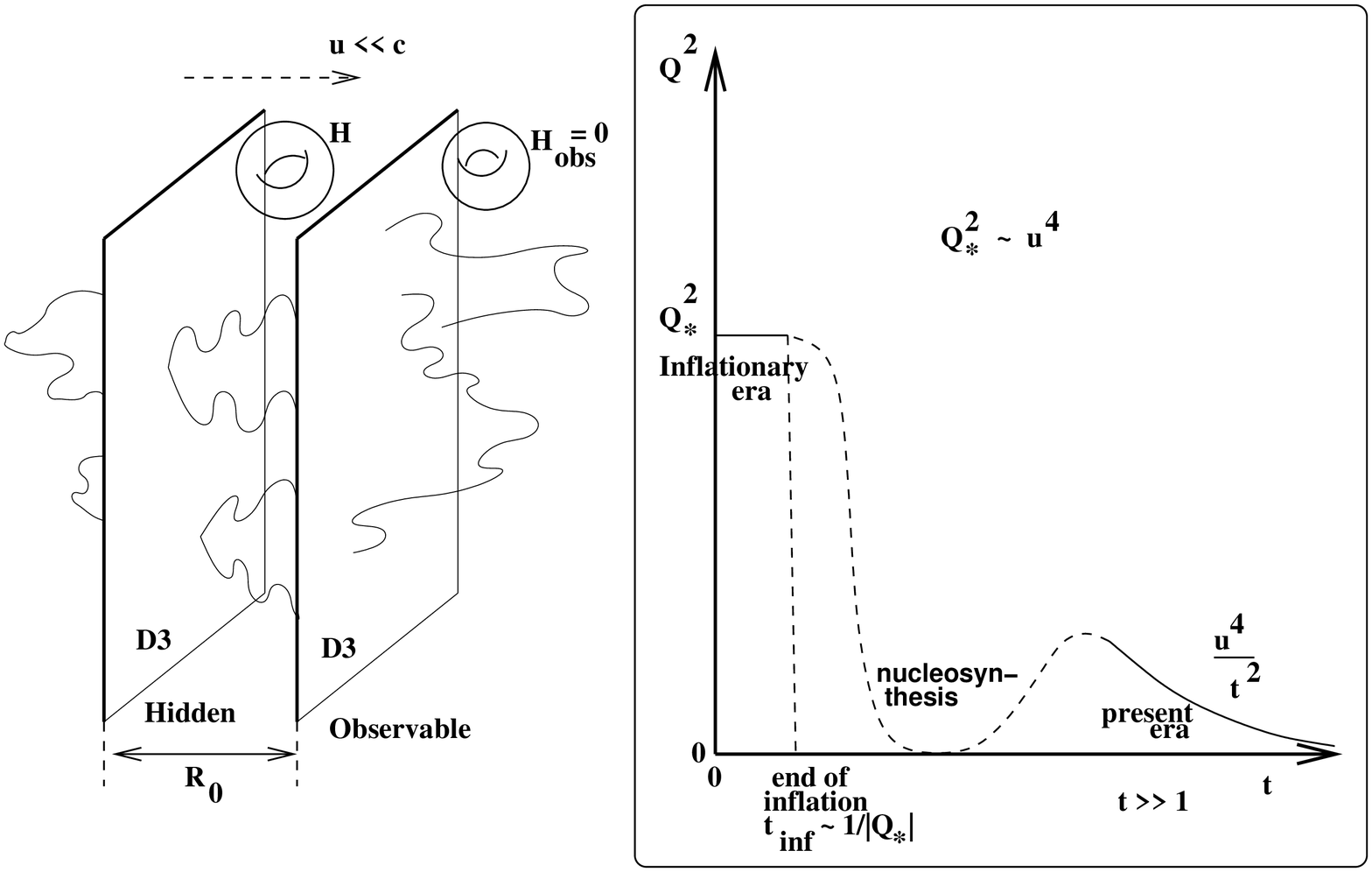}
\end{center} 
\caption[]{ A scenario 
involving two colliding type-II  5-branes which provides 
inflation and a 
relaxation model for cosmological vacuum energy.}
\label{fig:collbr} 
\end{figure}

The appearance of the magnetic field on the visible 
brane, on the dimensions $X^{4,5}$, 
is described (for times long after the collision)
within a $\sigma$-model
superstring formalism by the boundary deformation:
${\cal V}_{H}= \int _{\partial \Sigma} 
A_5\partial_\tau X^5 + {\rm supersymm.~partners}$,
where $A_5=e^{\varepsilon X^0} H X^4$, and 
$\partial_\tau$ denotes tangential $\sigma$-model derivative
on the world-sheet boundary.
This $\sigma$-model 
deformation 
describes open-string excitations attached to the brane world
(c.f. fig. \ref{fig:collbr}).
The presence of the quantity 
$\varepsilon \to 0^+$
reflects the {\it adiabatic} switching 
on of the magnetic field
after the collision. It should be remarked that in our approach 
the quantity $\varepsilon$ is viewed as a world-sheet 
renormalization-group scale parameter~\cite{gm}.  
In addition to the magnetic field deformation, the $\sigma$-model 
contains also boundary deformations describing the `recoil' of the 
visible world due to the collision:
${\cal V}_{\rm rec} = \int _{\partial \Sigma}  Y_6(X_0)\partial _n X^6
+{\rm supersymm.~partners}$,
where $Y_6(X_0)=u X^0 e^{\varepsilon X^0}$,
$\partial_n$ denotes normal $\sigma$-model derivative
on the world-sheet boundary, and $u$ is 
the recoil velocity of the visible brane world.

The presence of the exponential 
factors $e^{\varepsilon X^0}$ in 
{\it both} the magnetic field and recoil 
deformations 
implies a small but negative 
world-sheet anomalous dimension
$-\frac{\varepsilon ^2}{2} < 0$, and hence the relevance of 
both operators from a renormalization-group
point of view. The induced central-charge deficit $Q^2$,
which quantifies the departure from the conformal point of the 
pertinent $\sigma$-model, can be 
computed
by virtue of the Zamolodchikov's c-theorem~\cite{gm}:
$\frac{d}{d {\cal T}}Q^2 \sim - \frac{H^2 + u^2}{{\cal T}^2} \quad \to
\quad 
Q^2 ({\cal T}) = Q_0^2 + \frac{H^2 + u^2}{{\cal T}}~,~{\cal T}
={\rm ln}|L/a|^2~$, with $L (a)$ the 
world-sheet infrared (ultraviolet) cutoff scale.
As discussed in detail in 
\cite{gm}, the correct scaling behaviour
of the operators necessitates the 
identification ${\cal T} \sim \varepsilon^{-2}$ which we assume from now on. 
The quantity 
$Q_0^2 = Q^2 (\infty)$  
is the equilibrium
vacuum energy density, which we take 
to be zero $Q_0^2 =0$ due to an assumed cancellation between 
Anti-de-Sitter bulk and brane vacuum energies
after the collision.
We shall come back to this point later on.

The non-conformal deformed $\sigma$-model can become conformal
by Liouville dressing~\cite{ddk,gm}, as discussed in section 3:
$
A_5(X_0,X_4,\phi)=H X^4e^{\varepsilon X^0+ \alpha \phi} , \;\; 
Y_6(X_0,\phi)=u X^0 e^{\varepsilon X^0+ \alpha \phi}$,
where $\alpha \sim \varepsilon$ are the Liouville anomalous 
dimensions~\cite{ddk} and $\phi$ is the normalized Liouville mode,
whose zero-mode is related to the renormalization-group scale ${\cal T}$ 
as: $\phi = Q({\cal T}){\cal T}$. From 
the work of \cite{bachas} it becomes clear that the coupling constant 
$H$ is associated with supersymmetry-breaking mass splittings.
This has to do with the different way fermions and bosons couple to an 
external magnetic field. The mass splittings squared of an open string 
are in our case~\cite{gm}: 
\begin{equation} 
\Delta m_{\rm string}^2 \sim   He^{\alpha \phi + 
\varepsilon X^0 }\Sigma_{45} 
\label{split}
\end{equation}
The so-obtained mass splittings are constant upon the requirement 
that the flow of time $X^0$ and of Liouville mode $\phi$ are correlated
in such a way that 
\begin{equation} 
\varepsilon X^0 + \varepsilon \phi/\sqrt{2} 
={\rm constant}~,
\label{liouvtime}
\end{equation}
or at most slowly varying.
Notice that
deviations from the condition (\ref{liouvtime}) 
would result in very
large negative-mass squares, which are clearly unstable configurations.
Hence, the identification (\ref{liouvtime}) seems to provide
a resolution of this problem.
To ensure the phenomenologically reasonable order of magnitude of a TeV scale,
one must assume very small~\cite{bachas} $H \sim 10^{-30} \ll 1$ 
in Planck units.
Note also that parametrizing the condition (\ref{liouvtime})  
as $X^0=t$, $\phi_0 =\sqrt{2}t$,
and taking into account that, for convergence of $\sigma$-model 
path integration, it is formally necessary to work with Euclidean signature 
$X^0$~\cite{kmw}, 
the induced 
metric on the hypersurface (\ref{liouvtime}) in the 
extended space time  
acquires a 
Minkowskian-signature Robertson-Walker form: $ds^2_{\rm hypersurf} = -(d\phi_0)^2 + (dX^0)^2 + \dots = - (dt)^2 + \dots~$. 
where $\dots$ denote spatial parts. In \cite{gm}, where we refer the 
interested reader,
we have given some arguments on a 
a {\it dynamical stability} of the condition (\ref{liouvtime}) in the 
context of Liouville strings. 
Physically, one may interpret this 
result as 
implying that a time-varying magnetic field 
induced by the collision implies
back reaction of strings onto the space time 
in such a way that the mass splittings of the string excitation
spectrum, as a result of the field, are actually stabilized.

We now notice that in our case the dilaton field is 
$\Phi = Q\phi = Q^2 \varphi \sim (H^2 + u^2)$,
that is, one faces a situation with an asymptotically 
constant dilaton. 
This is a welcome fact, because otherwise, the space-time
would not be asymptotically flat, and one could face trouble in
appropriately defining masses. In the case of a constant dilaton the 
vacuum energy is determined by the central-charge charge 
deficit $Q^2$, which in our case is: 
\begin{equation}
\Lambda = \frac{R^{2n}}{\phi_0^2}(H^2 + u^2)^2
\label{cosmoconst}
\end{equation}
where $\phi_0=t$ is the world-sheet zero mode of the
Liouville field to be identified with the target time
on the {\it hypersurface} (\ref{liouvtime}) of the extended
space-time resulting after Liouville dressing. 

We next remark that the restoration of the conformal 
invariance by the Liouville mode results in the following equations
(c.f. (\ref{liouvcond})) 
for the $\sigma$-model background fields/couplings 
$g^i$ near a fixed-point of the world-sheet renormalization group
(large-times cosmology) we restrict ourselves 
here~\cite{ddk}:
\begin{equation} 
(g^i)'' + Q (g^i)' = -\beta^i (g)~,
\label{liouvcond2}
\end{equation}
where the prime denotes derivative with respect to the 
Liouville zero mode $\phi_0$, and the sign 
on the right-hand-side is appropriate for supercritical 
strings we are dealing with here. 
These equations replace the equilibrium equations $\beta^i=0$ 
of critical string theory, and should be used in our colliding brane scenario
to determine the evolution of the scale factor of the four-dimensional 
Robertson-Walker Universe. A preliminary analysis has been performed
in \cite{gm}, where we refer the reader for details.

Below we only describe briefly the main results. 
In our non-critical string scenario,
one does indeed obtain an expanding Universe, in contrast to 
standard ekpyrotic scenaria~\cite{ekpyrotic,linde}, based on 
critical strings and specific solutions to classical equations 
of motion. 
One of the most important features of the existence of a non-equilibrium 
phase of string theory due to the collision is the possibility
for an {\it inflationary phase}. Although the physics near the collision
is strongly coupled, and the $\sigma$-model perturbation theory
is not reliable, nevertheless one can give compelling 
physical arguments favoring the existence of an
early phase of the brane world where the four-dimensional 
Universe 
scale factor undergoes  exponential growth (inflation). 
This can be understood as follows: 
in our model we encounter two type-II string theory branes colliding, and then 
bouncing back. From a stringy point of view the collision and bounce 
will be described by a phase where open strings stretch between the 
two branes worlds (which can be thought of as lying a few string scales
apart during the collision, c.f. fig. \ref{fig:collbr}). During that early 
phase the excitation
energy of the brane worlds can be easily computed by the same methods
as those used to study scattering of type II D-branes in \cite{bachas2}.
In type II strings the exchange of pairs of open strings
is described by annulus world-sheet diagrams, which in turn 
results in the appearance of ``spin structure factors'' 
in the scattering amplitude. The latter are expressed in terms of 
appropriate sums over Jacobi $\Theta$ functions. 
Due to special properties of these functions, 
the 
spin structures start of at quartic order in $u$~\cite{bachas2}.
The resulting excitation energy is therefore 
of order ${\cal O}(u^4)$ and may be thought 
off as an initial value of the central charge deficit of the 
non-critical string theory describing the 
physics of our brane world after the collision. 
The deficit $Q^2$ 
is 
thus cut off at a finite value in the (world-sheet) 
infrared scale (early target times, c.f. fig. \ref{fig:collbr}). 
One may plausibly assume that the central charge deficit remains constant
for some time, which is the era of {\it inflation}. 
It can be shown~\cite{gm} that for (finite) 
constant 
$Q^2=Q_*^2= {\cal O}(u^4)$ the Liouville equations imply 
a scale factor exponentially growing with the Liouville zero mode  
$a(\phi_0) = e^{|Q_*|\phi_0/2}$
(assuming a negative $Q_*$). Upon the condition (\ref{liouvtime}), then, 
one obtains an early inflationary phase after the collision, in contrast 
to 
the critical-string based arguments of \cite{linde}. The duration of 
the inflationary phase is $t_{\rm inf} \sim 1/|Q_*| \sim {\cal O}(u^{-2})$,
which yields the conventional values of inflationary models
of order $t_{\rm infl} \sim 10^9 t_{\rm Planck}$ for $u^2 \sim 
10^{-9}$.

Exit from the inflationary phase is possible, and the de Sitter phase 
is assumed to be 
succeeded by a phase, to be identified with the 
{\it nucleosynthesis era},  in which the 
central charge deficit passes through a metastable critical phase
in which $Q$ vanishes, and then 
changes sign (cf. fig. \ref{fig:collbr}).
This can happen in Liouville strings with time-like fields $X^0$, 
as we mentioned above, where the central charge deficit can oscillate 
before it reaches its equilibrium value~\cite{schmid,diamand,gm}.
During the metastable critical phase the central charge deficit is zero
and hence there is no Liouville dressing. This implies that 
during that epoch any time dependence of fields 
on our brane world will be given by the ordinary critical string time
dependence. In brane cosmologies the Friedman equation on the 
three brane, where matter lives, assumes the form~\cite{langlois}
(for spatially flat branes):
\begin{equation}\label{branefried}
\left(\frac{\dot a}{a}\right)^2 = -\frac{|\Lambda_5|}{6} + \frac{\kappa^4}{36}\rho^2 + 
\frac{{\cal C}}{a^4}
\end{equation} 
where $\kappa$ is the gravitational constant in the bulk geometry, 
${\cal C}$ is an integration constant (a mass of a bulk Schwarzschild
black hole in the brane cosmology of \cite{langlois}, which 
for simplicity, and without any important physical consequences, we assume 
zero for our purposes),  
$\Lambda_5 < 0$ is the bulk anti de Sitter Cosmological constant,
and $a$ is the scale factor on the brane world. The density $\rho$ denotes
the total energy density on the brane, including matter contributions as well
as cosmological constant (brane tension) contributions. 
Shifting $\rho $ to make such tension contributions explicit, $\rho = c_0 + 
\rho_M$, where $\rho_M $ is the matter density, 
one observes that for late times, the Friedman equation on the brane assumes
the standard form, linear in $\rho_M$, {\it provided } one cancels
the total vacuum energy as perceived by a brane (four-dimensional) observer:
$c_0^2 -|\Lambda_5| =0$. In our non-critical string case $c_0$ may be 
assumed positive, and one may identify $Q_0^2=Q^2(\infty)$ 
(the equilibrium value of the 
string central charge deficit ) with $c_0^2 -|\Lambda_5| $ which vanishes.
During the nucleosynthesis era, therefore, the rate of expansion of the 
non-critical string 
Universe is that predicted by standard cosmology, which is a welcome feature
given the phenomenological 
success of the nucleosynthesis model in explaining 
the abundance of light elements in the Universe.

At the end of the nucleosynthesis one faces a non-critical 
situation, with a central charge deficit significantly smaller than that 
of the inflationary era. The above considerations on the two times,
Liouville $\phi$ and coordinate time $X^0$, arise again. 
The generalized conformal invariance conditions (\ref{liouvcond2}) 
apply in this case, and hence the scale factor depends now
on both $X^0$ and $\phi$, but the evolution 
is confined for energetic reasons on the hypersurface (\ref{liouvtime}),
as explained above.
We may assume that at the end of the nucleosynthesis $\beta^i=0$ for 
the scale factor, and hence one obtains from (\ref{liouvcond2})
that a natural solution is 
\begin{equation} 
a(\phi, X^0) \sim a_0(X^0)
\phi^{1/2 - {\cal O}(u^2,H^2)}~, 
\label{nucleo}
\end{equation} 
Since after the end of nucleosynthesis (present era) 
the universe is in a matter dominated
phase, one may assume, from the above considerations, that 
$a_0 \sim (t)^{2/3}$
with $X^0 = -t$ the Robertson-Walker observable time to be connected
with the Liouville time $\phi$ on the hypersurface (\ref{liouvtime}). 
Thus the present era scale factor scales with time as:
$a(t) \sim t^{2/3 + 1/2} = t^{7/6}$.
This suffices to yield a current acceleration of the universe, thereby 
indicating that the dark energy (in the form of gravitational recoil-Liouville
contributions) takes over the expansion. Unfortunately, 
such a scaling must not 
remain for ever, if one wants to recover an S-matrix equilibrium 
theory asymptotically (although it must be noted that 
this is not formally necessary in this framework, since Liouville
strings do not admit S-matrix description in general, as we have 
discussed above).

Remarkably a solution to this problem can be provided in our model
by noting the possibility of decompactification of the
extra toroidal dimensions on the five branes, as a result of the collision.
If one assumes a very slow decompactification 
(much slower than any other rate in the problem)
then eventually the radius of the torus 
$R \to \infty$, while keeping the magnetic 
field energy 
$(HR)^2$ constant. The masses of the matter particles on the brane will therefore 
eventually 
go to zero, and hence one will enter a radiation era in which 
$\rho_M \sim 1/a^6$ on the five brane. From (\ref{branefried}) 
this will imply an asymptotic  
scaling $a_0(t) \sim t^{1/3}$, which, upon Liouville dressing, will yield 
a scale factor $a \sim t^{1/3 + 1/2} \sim t^{5/6} $, as $t \to \infty$,
thereby resolving the cosmic horizon 
problem (\ref{cosmichorizon}),
and thus saving the S-matrix in this model. 
This concludes our discussion on this toy model.
It remains to be seen whether such conjectures/speculations  
can be accommodated
in realistic string models of brane cosmologies.

\section*{Acknowledgements}

\noindent I wish to thank Prof. H. Klapdor-Kleingrothaus and 
the organizers of the Conference
{\it Beyond the Desert 2002} (Oulu, Finland, June 2002) for the invitation.
This work is supported by the European Union (contract HPRN-CT-2000-00152).


\section*{References} 


\begin{thebibliography}{999}
\normalsize

\bibitem{uhecr} For a recent review see: 
P.~Biermann and G.~Sigl,
Lect.\ Notes Phys.\  {\bf 576} (2001) 1
[arXiv:astro-ph/0202425], and references therein

\bibitem{snae} S.~Perlmutter {\it et al.}  [Supernova Cosmology\
 Project Collaboration]: Nature {\bf 391}, 51 (1998);
A.~G.~Riess {\it et al.}  [Supernova Search Team Collaboration]:
Astron.\ J.\  {\bf 116}, 1009 (1998);
P.~M.~Garnavich {\it et al.}, Astrophys.\ J.\  {\bf 509}, 74 (1998).

\bibitem{cmb} J.R. Bond, A.H. Jaffe and L. Knox, Phys.\ Rev.\ D {\bf 57}, 
2117 (1998); H.~Lineweaver, astro-ph/9810334; N.~A.~Bahcall and X.~Fan,
astro-ph/9804082. 




\bibitem{glashow} for a subset of references see: 
S.~R.~Coleman and S.~L.~Glashow,
Phys.\ Rev.\ D {\bf 59} (1999) 116008
[arXiv:hep-ph/9812418].
S.~R.~Coleman and S.~L.~Glashow,
arXiv:hep-ph/9808446.
S.~R.~Coleman and S.~L.~Glashow,
Phys.\ Lett.\ B {\bf 405} (1997) 249
[arXiv:hep-ph/9703240]; 
O.~Bertolami,
Gen.\ Rel.\ Grav.\  {\bf 34} (2002) 707
[arXiv:astro-ph/0012462];
Nucl.\ Phys.\ Proc.\ Suppl.\  {\bf 88} (2000) 49
[arXiv:gr-qc/0001097].
O.~Bertolami and C.~S.~Carvalho,
Phys.\ Rev.\ D {\bf 61} (2000) 103002
[arXiv:gr-qc/9912117].




\bibitem{kifune} T.~Kifune,
Astrophys.\ J.\  {\bf 518} (1999) L21
[arXiv:astro-ph/9904164];
G.~Amelino-Camelia and T.~Piran,
Phys.\ Rev.\ D {\bf 64} (2001) 036005
[arXiv:astro-ph/0008107].
Phys.\ Rev.\ D {\bf 64} (2001) 036005
[arXiv:astro-ph/0008107];
Y.~J.~Ng, D.~S.~Lee, M.~C.~Oh and H.~van Dam,
Phys.\ Lett.\ B {\bf 507} (2001) 236
[arXiv:hep-ph/0010152].
J.~R.~Ellis, N.~E.~Mavromatos and D.~V.~Nanopoulos,
Phys.\ Rev.\ D {\bf 63} (2001) 124025
[arXiv:hep-th/0012216].


\bibitem{sarkar}
S.~Sarkar,
Mod.\ Phys.\ Lett.\ A {\bf 17} (2002) 1025
[arXiv:gr-qc/0204092], and references therein. 



\bibitem{gzk} K.~Greisen,
Phys.\ Rev.\ Lett.\  {\bf 16} (1966) 748;
G.~T.~Zatsepin and V.~A.~Kuzmin,
JETP Lett.\  {\bf 4} (1966) 78
[Pisma Zh.\ Eksp.\ Teor.\ Fiz.\  {\bf 4} (1966) 114].



\bibitem{challenge} 
T.~Banks, W.~Fischler: hep-th/0102077;
S.~Hellerman, N.~Kaloper and L.~Susskind,
JHEP {\bf 0106} (2001) 003
[arXiv:hep-th/0104180];
W.~Fischler, A.~Kashani-Poor, R.~McNees and S.~Paban,
JHEP {\bf 0107} (2001) 003
[arXiv:hep-th/0104181];
E. Witten, hep-th/0106109;
P.~O.~Mazur, E.~Mottola:
Phys.\ Rev.\ D {\bf 64}, 104022 (2001), and references therein.


\bibitem{protheroe} R.~J.~Protheroe and H.~Meyer,
Phys.\ Lett.\ B {\bf 493} (2000) 1
[arXiv:astro-ph/0005349].
See, however, K. Okumura {\it et al.}, astro-ph/0209487
which questions the experimental findings (thanks to A. Sakharov
for pointing this out). 


\bibitem{emn} J.~Ellis, N.~E.~Mavromatos, D.~V.~Nanopoulos: 
Phys.\ Lett.\ B {\bf 293} (1992) 37
[arXiv:hep-th/9207103];
Mod.\ Phys.\ Lett.\ A {\bf 10} (1995) 1685 
[arXiv:hep-th/9503162];
for reviews see: 
Erice Subnuclear Series (World Sci., Singapore) {\bf 31} 1, (1993);
[hep-th/9304133];
J. Chaos, Solitons and Fractals {\bf 10}, 345 (eds. C. Castro amd M.S. El Naschie,
Elsevier Science, Pergamon 1999) [hep-th/9805120], and references therein;
G.~Amelino-Camelia, J.~R.~Ellis, N.~E.~Mavromatos and D.~V.~Nanopoulos,
Int.\ J.\ Mod.\ Phys.\ A {\bf 12} (1997) 607
[arXiv:hep-th/9605211];
J.~R.~Ellis, N.~E.~Mavromatos and D.~V.~Nanopoulos,
Phys.\ Rev.\ D {\bf 62} (2000) 084019
[arXiv:gr-qc/0006004].
Phys.\ Rev.\ D {\bf 62} (2000) 084019
[arXiv:gr-qc/0006004];
E.~Gravanis and N.~E.~Mavromatos,
arXiv:hep-th/0103122.

\bibitem{garay} L.~Gonzalez-Mestres,
arXiv:physics/0003080;
L.~J.~Garay,
Phys.\ Rev.\ D {\bf 58} (1998) 124015
[arXiv:gr-qc/9806047].


\bibitem{nielsen} S.~Chadha and H.~B.~Nielsen,
Nucl.\ Phys.\ B {\bf 217} (1983) 125.
H.~B.~Nielsen and I.~Picek,
Nucl.\ Phys.\ B {\bf 211} (1983) 269
[Addendum-ibid.\ B {\bf 242} (1984) 542];
J.~L.~Chkareuli, C.~D.~Froggatt and H.~B.~Nielsen,
Phys.\ Rev.\ Lett.\  {\bf 87} (2001) 091601
[arXiv:hep-ph/0106036].

\bibitem{pascual} J.~I.~Latorre, P.~Pascual and R.~Tarrach,
Nucl.\ Phys.\ B {\bf 437} (1995) 60
[arXiv:hep-th/9408016].


\bibitem{nature} G.~Amelino-Camelia, J.~R.~Ellis, N.~E.~Mavromatos, D.~V.~Nanopoulos and S.~Sarkar,
Nature {\bf 393} (1998) 763
[arXiv:astro-ph/9712103];
J.~R.~Ellis, K.~Farakos, N.~E.~Mavromatos, V.~A.~Mitsou and D.~V.~Nanopoulos,
Astrophys.\ J.\  {\bf 535} (2000) 139
[arXiv:astro-ph/9907340];
J.~Ellis, N.~E.~Mavromatos, D.~V.~Nanopoulos and A.~S.~Sakharov,
arXiv:astro-ph/0210124.



\bibitem{pullin} R.~Gambini and J.~Pullin,
Phys.\ Rev.\ D {\bf 59} (1999) 124021
[arXiv:gr-qc/9809038];
arXiv:gr-qc/0110054.

\bibitem{ashtekar} A.~Ashtekar and J.~Lewandowski,
Class.\ Quant.\ Grav.\  {\bf 14} (1997) A55
[arXiv:gr-qc/9602046];
A.~Ashtekar, C.~Rovelli and L.~Smolin,
Phys.\ Rev.\ Lett.\  {\bf 69} (1992) 237
[arXiv:hep-th/9203079];
C.~Rovelli and L.~Smolin,
Phys.\ Rev.\ Lett.\  {\bf 61} (1988) 1155;
Nucl.\ Phys.\ B {\bf 331} (1990) 80;
For a comprehensive review (and more references) see: 
A.~Ashtekar,
arXiv:gr-qc/0112038.




\bibitem{biller} S.~D.~Biller {\it et al.},
Phys.\ Rev.\ Lett.\  {\bf 83} (1999) 2108
[arXiv:gr-qc/9810044].



\bibitem{amelino} G.~Amelino-Camelia,
Int.\ J.\ Mod.\ Phys.\ D {\bf 11} (2002) 35
[arXiv:gr-qc/0012051];
Nature {\bf 418} (2002) 34
[arXiv:gr-qc/0207049];
G.~Amelino-Camelia, D.~Benedetti and F.~D'Andrea,
arXiv:hep-th/0201245;
J.~Kowalski-Glikman,
Phys.\ Lett.\ A {\bf 286} (2001) 391
[arXiv:hep-th/0102098];
arXiv:hep-th/0209264.


\bibitem{smolin} J.~Magueijo and L.~Smolin,
Phys.\ Rev.\ Lett.\  {\bf 88} (2002) 190403
[arXiv:hep-th/0112090];
arXiv:gr-qc/0207085.


\bibitem{green}  
M.B. Green, J.H. Schwarz, E. Witten: \emph{Superstring  Theory}, 
Vols I and II (Cambridge University Press, Cambridge 1987).  

\bibitem{membranes} J.~Polchinski,
Phys.\ Rev.\ Lett.\  {\bf 75}, 4724 (1995);
{\it TASI lectures on D-branes}, hep-th/9611050;
M.~J.~Duff, Sci.\ Am.\  {\bf 278}, 64 (1998).


\bibitem{ddk} F.~David, Modern Physics Letters {\bf A3} (1988) 1651;
J.~Distler, H.~Kawai Nucl. Phys. {\bf B321} (1989) 509;
see also N.E. Mavromatos and J.L. Miramontes, Mod.
Phys. Lett. {\bf A4} (1989)  1847 (1989); E. D' Hoker and P.S. Kurzepa,
Mod. Phys. Lett. {\bf A5} (1990) 1411; E. D' Hoker, Mod. Phys. Lett. 
{\bf A6} (1991) 745.



\bibitem{zam} A.B. Zamolodchikov, 
JETP \ Lett.\ \textbf{43} (1986) 730. 


\bibitem{kutasov} D.~Kutasov, N.~Seiberg:
Nucl.\ Phys.\ B {\bf 358} (1991) 600.




\bibitem{schmid} C.~Schmidhuber, A.~A.~Tseytlin:
Nucl.\ Phys.\ B {\bf 426} (1994) 187.



\bibitem{aben} I.~Antoniadis, C.~Bachas, J.~R.~Ellis, D.~V.~Nanopoulos:
Phys.\ Lett.\ B {\bf 211} (1988) 393; 
Nucl.\ Phys.\ B {\bf 328} (1989) 117; Phys.\ Lett.\ B {\bf 257} (1991) 278.


\bibitem{diamand} G.~A.~Diamandis, B.~C.~Georgalas, 
N.~E.~Mavromatos, E.~Papantonopoulos and I.~Pappa,
Int.\ J.\ Mod.\ Phys.\ A {\bf 17} (2002) 2241
[arXiv:hep-th/0107124].

\bibitem{gm} E.~Gravanis and N.~E.~Mavromatos,
arXiv:hep-th/0205298, Physics Lett. {\bf B} in press, and work in progress.




\bibitem{ctp} The CTP formalism may be attributed to:
J. Schwinger, J. Math. Phys. {\bf 2}, 407 (1961);
for reviews see:
E. Calzetta and B.L. Hu, Phys. Rev. {\bf D37}, 2878 (1988);
E. Calzetta, S. Habib and B.L. Hu, Phys. Rev. {\bf D37}, 2901 (1988);
H. Umezawa,
{\it Advanced Field Theory: micro, macro and thermal concepts}
(American Inst. of Physics, N.Y. 1993).

\bibitem{hawk} S.~W.~Hawking, Commun.\ Math.\ Phys.\  {\bf 87}, 395 (1982).


\bibitem{kogan1} Y.~I.~Kogan,
Phys.\ Lett.\ B {\bf 265} (1991) 269.


\bibitem{kogan2} Y.~I.~Kogan, Univ. of British Columbia preprint 
UBCTP 91-13 (1991).

\bibitem{szabo} A non-linear difusive non-critical stringy 
system 
is examined in: N.~E.~Mavromatos and R.~J.~Szabo,
Int.\ J.\ Mod.\ Phys.\ A {\bf 16} (2001) 209
[arXiv:hep-th/9909129];
see, however, how world-sheet supersymmetry may eliminate such non-linearities:
N.~E.~Mavromatos and R.~J.~Szabo,
JHEP {\bf 0110} (2001) 027
[arXiv:hep-th/0106259].


\bibitem{kmw} I.~I.~Kogan, N.~E.~Mavromatos and J.~F.~Wheater,
Phys.\ Lett.\ B {\bf 387} (1996) 483
[arXiv:hep-th/9606102];
N.~E.~Mavromatos and R.~J.~Szabo,
Phys.\ Rev.\ D {\bf 59} (1999) 104018
[arXiv:hep-th/9808124].

\bibitem{egmn} J.~Ellis, E.~Gravanis, N.~E.~Mavromatos and D.~V.~Nanopoulos,
arXiv:gr-qc/0209108.


\bibitem{SUV} D.~Sudarsky, L.~Urrutia and H.~Vucetich,
arXiv:gr-qc/0204027;
J.~Alfaro, H.~A.~Morales-Tecotl and L.~F.~Urrutia,
Phys.\ Rev.\ Lett.\  {\bf 84}, 2318 (2000)
[arXiv:gr-qc/9909079];
J.~Alfaro and G.~Palma,
Phys.\ Rev.\ D {\bf 65}, 103516 (2002)
[arXiv:hep-th/0111176].





\bibitem{carroll} For a review see: 
S.~M.~Carroll, Living Rev.\ Rel.\  {\bf 4}, 1 (2001) [astro-ph/0004075].










\bibitem{emnaccel} J.~Ellis, N.~E.~Mavromatos, D.~V.~Nanopoulos: 
hep-th/0105206. 


\bibitem{bachas} C.~Bachas,
arXiv:hep-th/9503030.


\bibitem{ekpyrotic} J.~Khoury, B.~A.~Ovrut, P.~J.~Steinhardt and N.~Turok,
Phys.\ Rev.\ D {\bf 64}, 123522 (2001).

\bibitem{linde} For a recent 
and concise review of the criticisms of the model of 
\cite{ekpyrotic} see: A.~Linde,
arXiv:hep-th/0205259, and references therein. 

\bibitem{bachas2} C.~Bachas,
Phys.\ Lett.\ B {\bf 374}, 37 (1996)
[arXiv:hep-th/9511043].

\bibitem{langlois} For a recent review see:
D.~Langlois,
arXiv:hep-th/0209261 and references therein; 
see also: P.~Binetruy, C.~Deffayet and D.~Langlois,
Nucl.\ Phys.\ B {\bf 565} (2000) 269
[arXiv:hep-th/9905012];
P.~Binetruy, C.~Deffayet, U.~Ellwanger and D.~Langlois,
Phys.\ Lett.\ B {\bf 477} (2000) 285
[arXiv:hep-th/9910219].

\end{thebibliography}
\end{document}